\documentclass[%
superscriptaddress,
twocolumn,
 amsmath,amssymb,
 aps,
prb,
floatfix,
]{revtex4-2}

\usepackage{graphicx}
\usepackage{dcolumn}
\usepackage{bm}
\usepackage{upgreek}
\usepackage{xcolor}
\usepackage{hyperref}

\newcommand{\FBO}{Fe$_3$BO$_6$}

\newcommand{\Tsr}{$T_\mathrm{SR}$}
\newcommand{\Gtwo}{$\Gamma_2$}
\newcommand{\Gfour}{$\Gamma_4$}
\newcommand{\TO}{$T_0$}

\newcommand{\TT}{$\uptau_\mathrm{L}$}
\newcommand{\tsr}{$t_\mathrm{SR}$}

\definecolor{AMK}{RGB}{0, 0, 250}
\definecolor{Anna}{RGB}{250, 0, 0}
\begin{document}

\preprint{APS/123-QED}

\title{
Revealing laser-induced modification of weak ferromagnetism\\ through a spin-reorientation phase transition in iron borate \FBO}
\author{A.~V. Kuzikova}
\email{anna.kuzikova@mail.ioffe.ru}
\affiliation{%
 Ioffe Institute, 194021 St.~Petersburg, Russia}
\author{S.~N. Barilo}
\affiliation{%
Scientific-Practical Materials Research Centre, NAS of Belarus, 
220072 Minsk, Belarus}
\author{R.~A.~Lanovsky}
\affiliation{%
Scientific-Practical Materials Research Centre, NAS of Belarus, 
220072 Minsk, Belarus}
\author{A.~V. Sadovnikov}
\affiliation{%
Laboratory ”Magnetic Metamaterials”, Saratov State University, 41001 Saratov, Russia}
\author{R.~V. Pisarev}
\affiliation{%
 Ioffe Institute, 194021 St.~Petersburg, Russia}
\author{A.~M. Kalashnikova}
\affiliation{%
 Ioffe Institute, 194021 St.~Petersburg, Russia}

\date{\today}

\begin{abstract}
In the iron oxide \FBO{} weak ferromagnetism stems from competing contributions from two-site Dzyaloshinskii-Moriya interactions (DMI) and from on-site anisotropies at the two $4c$ and $8d$ types of nonequivalent Fe$^{3+}$ sites. 
As a result, spin reorientation (SR) transition \Gtwo{} $\rightarrow$ \Gfour{} in \FBO{} uniquely comprises 90$^\circ$ spin switching and a 35\% change of weak ferromagnetic moment. 
We explore ultrafast kinetics of such a transition as a proxy for weak ferromagnetism contributions modified by femtosecond laser pulses. 
Using magneto-optical pump-probe technique we reveal that the SR transition is triggered by an abrupt decrease of the weak ferromagnetic moment in the \Gtwo{} phase, and relate it to the laser-induced modification of the DMI and anisotropy contributions to the spin canting.
By tuning the pump pulse phonon energy between different optical absorption ranges, we show that laser-driven impact on spin canting possesses a nonthermal character and is achieved through the charge-transfer transitions $2p\rightarrow3d$ within Fe$^{3+}$O$^{2-}_6$ octahedra rather than by localized $3d\rightarrow3d$ transitions at Fe$^{3+}$.
The decreased weak ferromagnetic moment controls the threshold and the latency of the SR transition, as well as a final state.
The latter is achieved via the precessional switching to the \Gfour{} phase activated by the laser-induced heating.
Final state is found to be a coexistence of \Gfour{} and \Gtwo{} phase with modified spin canting persisting even at hundreds of picoseconds. 
\end{abstract}

\maketitle
\section{\label{sec:intro}Introduction}

The kinetics of phase transitions reveals how materials transform from one state to another, determining the pathways, timescales, and intermediate states that emerge during such transformations. 
The 
ultrafast laser techniques has enabled the exploration of phase transitions kinetics far from equilibrium, where the system is driven on timescales comparable to or shorter than the relaxation times of its degrees of freedom \cite{Koshihara-PhysRep2022, Fausti2011, de2013speed, WEGKAMP2015, li2022ultrafast, de2012coherent, Kurihara-PRL2018}. 
Laser excitation can induce transient, hidden, and metastable phases inaccessible under equilibrium conditions, and thus alter the very pathways by which a system transforms. 
Phase transitions in magnetic materials, with their coupled spin, lattice, and electronic subsystems, provide a particularly interesting and challenging task for such studies \cite{ichikawa2011transient,li2026discovery, mariager2012structural, radu2011transient, kimel2004laser}.

Spin-reorientation (SR) transitions being the abrupt switching of the magnetic moments from one crystallographic axis to another represent a particularly interesting case for studies under ultrafast laser excitation.
Indeed, they rely on a delicate balance between several interactions stabilizing a particular spin-orientational states \cite{belov1976spin}.
Diverse microscopic sources of magnetic anisotropy provide such a balance in various crystals and structures.
These sources include include spin-orbital coupling giving rise to a conventional single-ion as well as 
two-ion \cite{Mryasov2005twoion} contributions, exchange coupling exploited for a unidirectional anisotropy \cite{stamps2000mechanisms}, two-ion spin-orbital coupling yielding Dzyaloshinskii-Moriya interaction (DMI) \cite{Moskvin-CondMat2019,Mazurenko-JETP2021}, as well as many-ion magnetic-dipolar interactions playing a crucial role in high-magnetization materials and nanostructures. 
Changing either of these interactions may steer the material through SR transition.
Thus, studies of SR transitions kinetics in both laser-driven and equilibrium regimes are indispensable for a profound understanding of various contributions to magnetic anisotropy.
Examples include ultrafast laser heating as a universal approach to both intrinsic \cite{kimel2004laser} and engineered \cite{shelukhin2022spin} SR transitions; revealing a link between finite timescale of a SR transition and electron-lattice relaxation in rare-earth orthoferrites \cite{de2011laser}, exploiting nonthermal pathways to control anisotropy via laser-driven phonons \cite{afanasiev2021ultrafast} and THz electric fields \cite{schlauderer2019temporal}, unraveling spin fluctuations in a vicinity of SR transitions \cite{weiss2023discovery}, and many others.

In antiferromagnets, interactions favoring spin canting, i.e. weak ferromagnetism \cite{dzyaloshinsky1958thermodynamic}, play a significant role in emerging of SR transitions, as this phenomenon is genuinely linked to orientational spin configurations \cite{moriya1960anisotropic}.
A textbook case is a SR transitions (Morin transition) in the hematite that also is a switching between antiferromagnetic and weak ferromagnetic states \cite{morin1950magnetic}.
Since laser-driven change and control of weak ferromagnetism is a subject of ongoing research nowadays \cite{mikhaylovskiy2015ultrafast,mikhaylovskiy2020resonant,Kuppevelt-APL2026,radovskaia2026photoengineering}, corresponding SR transitions could provide valuable insights into this problem.
In particular, materials with an intrinsic link between weak ferromagnetism and SR transition should facilitate resolving an absolute sign of laser-induced spin canting, 
as well as distinguishing between changes of antisymmetric and symmetric exchange couplings and anisotropy, which competition gives rise to weak ferromagnetism.

A canted antiferromagnet iron borate \FBO{} is a model material for such studies.
This compound takes a unique place among weak ferromagnets, with the spin canting having a twofold origin, namely, conventional two-site DMI and single-ion anisotropies at nonequivalent crystallographic sites of magnetic ions.
In equilibrium, the first-order SR transition in the iron borate involves a 90$^\circ$ reorientation of the magnetization, accompanied by a 35\% change of the latter within a narrow temperature interval of $\sim$1~K around \Tsr~=~415~K. 
This, on the one hand, raises a question regarding the leading parameter governing phase transition. 
On the other hand, it offers an opportunity to address the laser-driven changes of competing contributions to the spin canting.

In this Article, we report on the laser-driven kinetics of the SR transition in \FBO.
By performing the time-resolved magneto-optical pump-probe experiments, we demonstrate that a decisive role in driving the laser-induced SR transition is played by a reduction of the magnetization, achieved through modification of the spin canting. 
The SR transition is found to proceed via a hidden phase whose magnetization value corresponds to neither of the equilibrium states. 
We establish how the net magnetization in this hidden phase either hinders or enables complete 90° reorientation facilitated by laser-induced heating.
As for the laser-induced spin canting change, we reveal that it relies on laser-driven charge transfer transitions promoting electrons from 2$p$ O$^{2-}$ to 3$d$ Fe$^{3+}$  states thus altering both DMI and single-ion contributions non-thermally.
This process can be realized either by resonantly pumping the change-transfer transition or by pumping below the absorption edge and employing the two-photon absorption.
These results complement and extend the findings reported in Ref.~\cite{Kuzikova-PRL2025}.

This paper is organized as follows. 
In Sec.~\ref{sec:FBOoverview} we discuss the main properties of \FBO{}, origin of the weak ferromagnetism and characteristics of the equilibrium SR transition, as well as briefly overview its optical properties relevant to the present study. 
Section~\ref{sec:sample} introduces the samples and their characterization and outlines the experimental strategy to study the kinetics of laser-induced transition. 
In Sec.~\ref{sec:results} we discuss the experimental results and reveal features of laser-induced SR transition in \FBO{} by varying laser pulse fluence and photon energy, sample temperature, and applied magnetic field. 
In Sec.~\ref{sec:discution} we discuss signatures, non-thermal origin and possible mechanism underlying the laser-induced spin canting change in \FBO, analyze how it impacts the SR transition boundaries, and outline the complete scenario of the switching.
Section~\ref{sec:Conclusion} presents the concluding remarks and outlook.

\section{\label{sec:FBOoverview} Magnetic and optical properties of F\lowercase{e}$_3$BO$_6$}

\subsection{\label{sec:SR} Magnetic structure and spin-reorientation transition in \FBO{}}

\begin{figure}
	\includegraphics [width=0.47\textwidth] {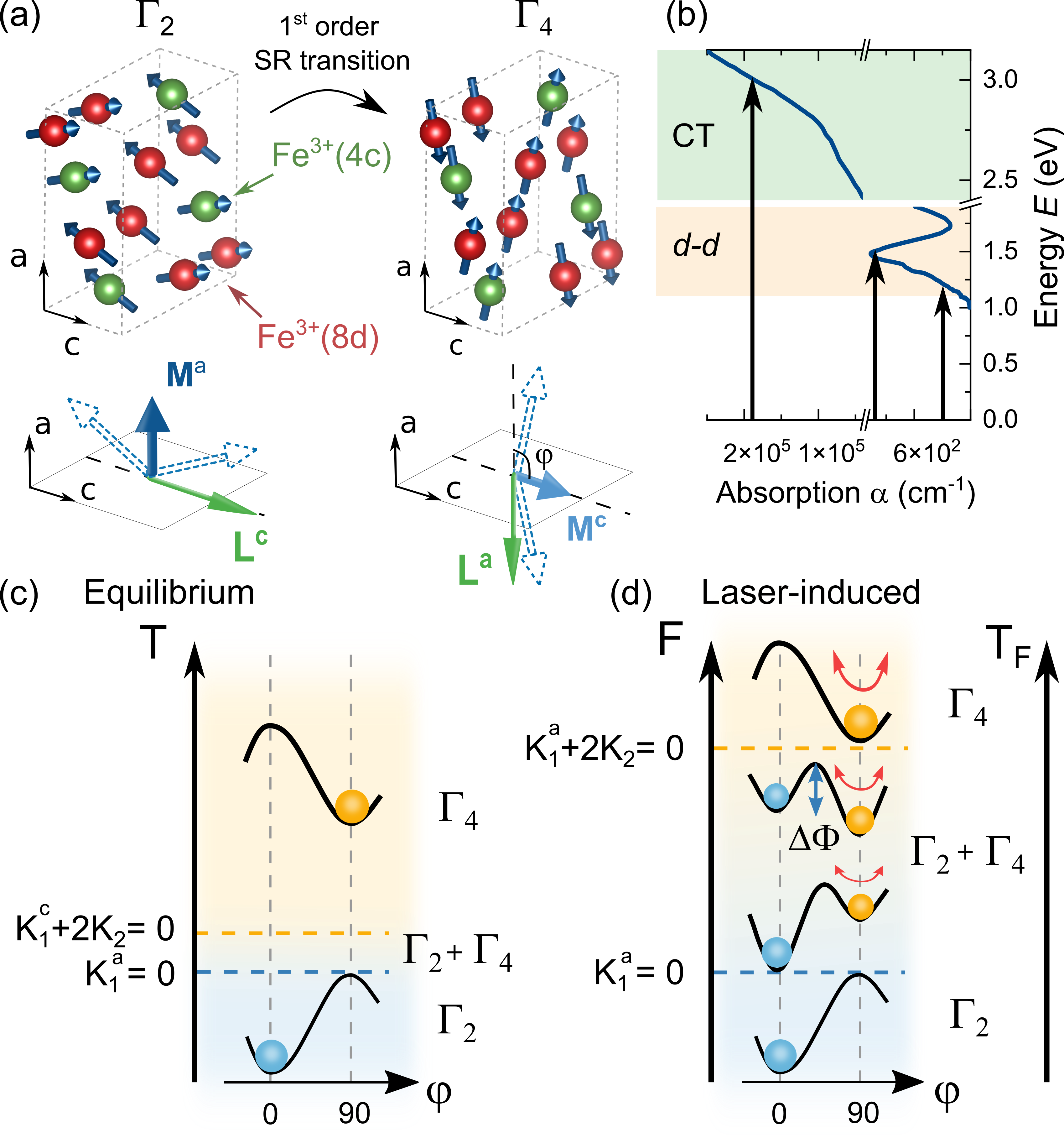}
	\caption{\label{fig:structure} 
		    (a) Crystal structure of \FBO, spins arrangement, net magnetization $\mathbf{M}$ and antiferromagnetic vector $\mathbf{L}$ below and above the SR transition.
		(b) Schematic absorption spectrum in a range 1-3~eV for a light polarized along the $c$-axis \cite{pisarev2009charge}.
        Vertical arrows indicate the excitation by the pump pulses with photon energy in a range of low and moderate absorption governed by the localized $d-d$ transitions (yellow band) and in the range of strong absorption due to CT transitions (green band).
        (c) Magnetic energy profiles $\Upphi(\upphi)$ at different equilibrium temperatures $T_0$ and corresponding phases: \Gtwo{} (blue background), \Gtwo+\Gfour (gradient background), and \Gfour{} (yellow background).
        Also shown are the conditions for the equilibrium phase boundaries [Eq.~\eqref{eq:bound}], where $K_{1}^{a(c)}$ depends on equilibrium $m^{a(c)}$ according to Eqs.~\eqref{eq:K12},~\eqref{eq:M}.    
        (d) Evolution of the magnetic energy profile at different laser excitation fluences $F$.
        In contrast to the equilibrium case, phase boundaries are defined by $K_{1}^{a}$ with $m^{a}$ reduced by laser-driven change of the spin canting.}
\end{figure}
The iron borate \FBO{} belongs to an orthorhombic space group $Pnma$ (crystallographic point group $mmm$) in which twelve Fe$^{3+}$ ions in the unit cell occupy two non-equivalent $8d$ and $4c$ distorted octahedral positions \cite{voigt1975temperature}.
Unit cell parameters are $a$ = 10.05, $b$ = 8.55 and $c$ = 4.47~$\dot{A}$~\cite{white1965fe3bo6}.
Magnetic moments of the iron ions within both positions order at the same temperature $T_N=508$~K. 
Within each position they are coupled antiferromagnetically and experience canting leading to a weak ferromagnetism, as illustrated in Fig.~\ref{fig:structure}(a).

Weak ferromagnetism in Fe$_3$BO$_6$ comprises a peculiar case when two mechanisms of the phenomenon, namely single-ion magnetic anisotropy and two-ion DMI, coexist and are of comparable strength.
We consider a two-sublattice model with the two magnetizations $\mathbf{M}_{1,2}=M_0\mathbf{m}_{1,2}$, where $\mathbf{m}_{1,2}$ are unit vectors.
Each magnetic sublattice consists of collinear magnetic moments at $8d$ and $4c$ positions in a proportion 2:1. 
Then ferromagnetic and antiferromagnetic (N\'eel) vectors are $\mathbf{M}=M_0\mathbf{m}$ and $\mathbf{L}=M_0\mathbf{l}$, respectively [Fig.~\ref{fig:structure}(a)], with
\begin{eqnarray}\mathbf{m}&=&0.5(\mathbf{m}_1+\mathbf{m}_2)\nonumber\\
	\mathbf{l}&=&0.5(\mathbf{m}_1-\mathbf{m}_2).
\end{eqnarray}
Phenomenologically, we write the thermodynamic invariant following \cite{ozhogin1972statics}:
\begin{eqnarray}
	\Upphi & = & \frac{E}{2} m^ 2 - d_a m_a l_c + d_c m_c l_a  \nonumber\\
	& + & \frac{a_1}{2} l^2 _c + \frac{c_1}{2} l^2 _a + \frac{a_2}{4} l^4 _c + \frac{c_2}{4} l^4 _a + \frac{f}{2} l^2 _c l^2 _a,
	\label{eq:DMI}
\end{eqnarray}
with $m_{a(c)}$ and $l_{a(c)}$ being the projections of $\mathbf{m}$ and $\mathbf{l}$ on the $a(c)-$axis. 
Here, $E/2>0$ is an effective exchange parameter describing antiferromagentic coupling between the sublattices, $a_1$, $c_1$, $a_2$, $c_2$, $f$ are the bilinear and biquadratic magnetocrystalline anisotropy parameters.
A pair of the parameters  $d_a$ and $d_c$ describes weak ferromagnetism, as the term $- d_a m_a l_c + d_c m_c l_a$ allows the non-zero net moment along either $a$- or $c$-axis, respectively.

The physical meaning of the distinct parameters $d_{a}$ and $d_{c}$ is evident if one rewrites the relevant term in Eq.~\eqref{eq:DMI} in the form with sublattice magnetizations \cite{Turov-book,Moskvin-CondMat2019}:
\begin{align}
	D\mathbf{u}_b\cdot\left[\mathbf{m}_1\times\mathbf{m}_2\right]+d\left(m_{1a}m_{1c}-m_{2a}m_{2c}\right).\label{eq:Dzy}
\end{align}
Then $d_a=D+d$, $d_c=D-d$. 
Here $D\mathbf{u}_b$ is the Dzyaloshinskii-Moriya vector, which is aligned with the $b$-axis in the considered compound. 
The corresponding term describes the weak ferromagnetism originating from the DMI, with the sublattices canted away from the strictly antiparallel alignment by an angle $\propto DE^{-1}$.
The second term is essentially the single-ion-like contribution to the weak ferromagnetism enabled by different biaxial anisotropy at 4$b$ and 8$d$ sites.
Indeed, such a term favors $\mathbf{m}_1$ and $\mathbf{m}_2$ to orient orthogonally to each other in the $ac$-plane. 
Competing with the antiferromagnetic exchange coupling, this leads to the canting of sublattices by $\propto dE^{-1}$.

Significance of the both contributions to the weak ferromagnetism in \FBO{} is evident from its peculiar SR phase transition.
The transition $\Gamma_2\rightarrow\Gamma_4$ consists in an abrupt rotation of $\mathbf{m}$($\mathbf{l}$) in the $ac$-plane from $a(c)$-axis to $c(a)$-axis.
Importantly, the weak ferromagnetic moment changes both its orientation and magnitude upon the SR transition.
Indeed, from Eqs.~(\ref{eq:DMI}),(\ref{eq:Dzy}) it follows that the two contributions to the weak ferromagnetism either add to or compete with each other depending on the spin alignment within the $ac$-plane, since only the single-ion-like term changes its sign upon the 90$^\circ$ spin reorientation in this plane. 
This yields the orientational dependence of the total weak ferromagnetic moment:
\begin{eqnarray}
	m^a =\frac{d_a}{E-c_2} = \frac{D+d}{E-c_2}; \nonumber\\
	m^c =\frac{d_c}{E-a_2} = \frac{D-d}{E-a_2}, \label{eq:M}
\end{eqnarray}
where $m^{a(c)}$ denotes the magnitude of $\mathbf{m}$ when it is aligned along the $a(c)-$axis ~\cite{martynov2018single,turov1965physical}.
It is established that the SR transition in \FBO{} occurs at \Tsr{} = 415~K \cite{kamzin1993mossbauer, tsymbal2006orientation}.
The weak magnetic moment reorientation by 90$^\circ$ from the $a$ to the $c$-axis is accompanied by a reduction in its magnitude to $M_S^c$ = 0.6 $M_S^a$ \cite{tsymbal2006orientation}, where $M_S^{c(a)}=M_0m^{c(a)}$.
The pronounced difference in the magnetization in the $\Gamma_2$ and $\Gamma_4$ phases signifies that $d_a\neq d_c$ and the single-ion-like contribution to weak ferromagnetism ($d$) is comparable to the one related to the DMI ($D$).
A rough estimate using Eq.~\eqref{eq:M} and neglecting a difference in magnetic anisotropy parameters $a_2,c_a$ gives $d\approx0.25D$ in the vicinity of the SR transition.

Another distinct feature of the SR transition in \FBO{} is that it is of the first order and has a very narrow temperature hysteresis in a very narrow temperature interval of 0.5$-$3.0~K \cite{kamzin1993mossbauer,tsymbal2006orientation}.
To highlight the relevance of the weak ferromagnetism to this feature of the SR transition we use the conventional description of the transition based on an expansion of the thermodynamic potential \cite{belov1976spin}:
\begin{align}
\Phi  =  K_1\cos^2{\upvarphi} + K_2 \cos^4{\upvarphi},\label{eq:energy}
\end{align}
where $K_1$ and $K_2$ are effective anisotropy parameters, $\upvarphi$ is an angle between the weak ferrommagnetic moment $\mathbf{m}$ and the $a$-axis [Fig.~\ref{fig:structure}(a)]. 
The effective anisotropy constants $K_1$, $K_2$ in Eq.~\eqref{eq:energy} can be expressed via the parameters entering Eq.~\eqref{eq:DMI}:
\begin{align}
K_1 & =d_c m - d_a m 
- \frac{a_1}{2} + \frac{c_1}{2} - \frac{a_2}{2} + \frac{f}{2} \nonumber  \\ 
&=-2d m 
- \frac{a_1}{2} + \frac{c_1}{2} - \frac{a_2}{2} + \frac{f}{2}; \nonumber\\
K_2 & = 
\frac{a_2}{4} + \frac{c_2}{4} - \frac{f}{2},
\label{eq:K12}
\end{align}   
where conditions $l^2+m^2=$~1 and $m\ll$~1 were used, and only the terms linear on $m$ are preserved.
The first order character of the SR transition in \FBO{} sets the constrain $K_2<0$. 
The phase is given by the following relations
\begin{align}
    K_1<0 & :  &\Gamma_2  \nonumber\\
    K_1>0&, K_1+2K_2<0: & \Gamma_2+\Gamma_4 \label{eq:bound}\\
    &\, K_1+2K_2>0: & \Gamma_4 \nonumber
\end{align}
Corresponding energy profiles are shown in Fig.~\ref{fig:structure}(c).
The boundaries of the phase transition and phase coexistence are determined by both parameters responsible for the weak ferromagnetism in addition to the magnetocrystalline anisotropy. 
Narrowness of the transition corresponds to the two conditions for the phase boundaries satisfied at nearly the same temperature, $K_1(m^a)=0$ and 
$K_1(m^c)+2K_2=0$, that is enabled by $m^a\neq m^c$ and a particular ratio between the weak ferromagnetism parameters $D$ and $d$.

Whereas the DMI term in Eq.~\eqref{eq:Dzy} has been extensively studied over the decades and its origin from exchange-spin-orbital coupling is established \cite{Moskvin-CondMat2019,Mazurenko-JETP2021}, the single-ion-like term was investigated in detail only in a few antiferromagnets. 
This is despite the fact that the seminal works that introduced the phenomenon of weak ferromagnetism \cite{Dzyaloshinskiy-JPCS1958,Dzyaloshinskiy-JETP1958-fluorides,Moriya-PR1960,Moriya-PR1960-NiF2} already addressed the contribution of the anisotropy.. 
Weak ferromagnetism arising exclusively from the single-ion-like contribution was found experimentally in \textit{e.g.} NiF$_2$ \cite{Matarrese-PRB1954}
and some other materials \cite{petrakovskiui2009magnetic}, and was described theoretically in \cite{Dzyaloshinskiy-JETP1958-fluorides,Moriya-PR1960-NiF2,martynov2018single}. 
Recently, owing to an altermagnetism of NiF$_2$ \cite{Jo-PRL2025} the theory behind this contribution was revisited to emphasize a role of the nonequivalent crystallographic positions of magnetic Ni$^{2+}$ ions \cite{Jo-PRL2025}. 
It is argued that the corresponding $g$-tensor anisotropy yields significant orbital contribution to the weak magnetic moment \cite{Jo-PRL2025}.
Although the microscopic theory of the weak ferromagnetism in \FBO{} has not yet been addressed, the nonequivalent 8$d$ and 4$c$ positions occupied by Fe$^{3+}$ ions [Fig.~\ref{fig:structure}(a)] appear to be essential for a sizable magnitude of $d$.

Non-negligible single-ion contribution to the weak ferromagnetism as compared to the DMI distinguishes \FBO{} from other Fe$^{3+}$-based weak ferromagnets such as rare-earth (RE) orthoferrites $RE$FeO$_3$, hematite $\alpha$-Fe$_2$O$_3$, and another iron borate FeBO$_3$, where DMI is the leading interaction responsible for the spin canting.
In $\alpha$-Fe$_2$O$_3$ and FeBO$_3$, $d\equiv0$ as the single-ion-like contribution is not allowed by their symmetry \cite{Dzyaloshinskiy-JPCS1958}.
The SR (Morin) transition in $\alpha$-Fe$_2$O$_3$ involves switching between phases with symmetry-allowed and forbidden weak ferromagnetism \cite{morin1950magnetic}.
Although orthoferrites are described by the thermodynamic potential [Eq.~\eqref{eq:DMI}], it was found that $d<<D$ and the corresponding term can be neglected \cite{treves1962magnetic,ozhogin1972statics}.
Consequently, SR transitions of a $\Gamma_2\rightarrow\Gamma_4$ type observed in many orthoferrites result from peculiarities of the magnetocrystalline anisotropy related to the presence of RE ions, and preserve the magnitude of a weak magnetic moment\cite{belov1976spin}.

Such a distinct feature of the equilibrium SR transition in \FBO{} makes it a peculiar case for the study under ultrafast laser excitation.
Indeed, revealing transition kinetics should allow to distinguish and analyze stages through which the spin reorientation and change of a weak ferromagnetic moment occur, and to identify the leading factor in the whole transition.
Next, it should enable us to reveal and discern laser-induced changes of the two competing weak ferromagetism parameters $D$ and $d$. 
On the other hand, high temperature of the transition and its narrowness suggest that laser excitation offers a way to effectively switch the antiferromagnetic vector of this material by 90$^\circ$.

\subsection{\label{sec:opt} Optical properties of \FBO{}}

\FBO{} is a dielectric material with band gap above 2.5~eV \cite{pisarev2009charge}.
Below-bandgap absorption spectrum consists of $3d$-$3d$ transitions at Fe$^{3+}$ sites: $^6A_1 \rightarrow ^4T_1$ at 1.5~eV and $^6A_1 \rightarrow ^4T_2$ at 1.9~eV \cite{andlauer1977optical}.
Third band $^6A_1 \rightarrow ^4A_1,^4E$ at around 2.5~eV is at the tail of a electric-dipole-allowed bands responsible for fundamental absorption band gap \cite{pisarev2009charge}.
Pronounced distortions of the oxygen octahedra surrounding Fe$^{3+}$ at the both positions lead to increased oscillator strength of the $d-d$ transitions as compared to some other iron oxides.
Joint theoretical and epxerimental study of fundamental absorption bands in \FBO{} and a number of related iron oxides was reported in \cite{pisarev2009charge}.
It was shown that the absorption around 3.0~eV is dominated by electric-dipole-allowed $2p-3d$ charge-transfer (CT) transitions $t_{2u}(\pi)\rightarrow t_{2g}$ in FeO$_6$ octahedra \cite{pisarev2009charge}.
In both optical ranges, \FBO{} demonstrates noticeable trichroism \cite{andlauer1977optical,pisarev2009charge} closely related to its crystal structure.

In this work, we analyze the magnetization dynamics driven by laser pulses with central photon energies below and in the $d-d$ and CT transitions ranges.
Figure~\ref{fig:structure}(b) schematically reproduces relevant absorption spectra in the range of $^6A_1 \rightarrow ^4T_1$ transition from Ref.~\cite{andlauer1977optical} and of the fundamental absorption around 3.0~eV from Ref.~\cite{pisarev2009charge} for light polarized along the $c$-axis, that is in accordance with the pump pulse polarization as described in Sec.~\ref{sec:Pump-probe experiment}.

\section{\label{sec:sample} Sample and experimental details}
\subsection{\label{sec:smpl} Samples}
\begin{figure}
	\includegraphics [width=0.45\textwidth] {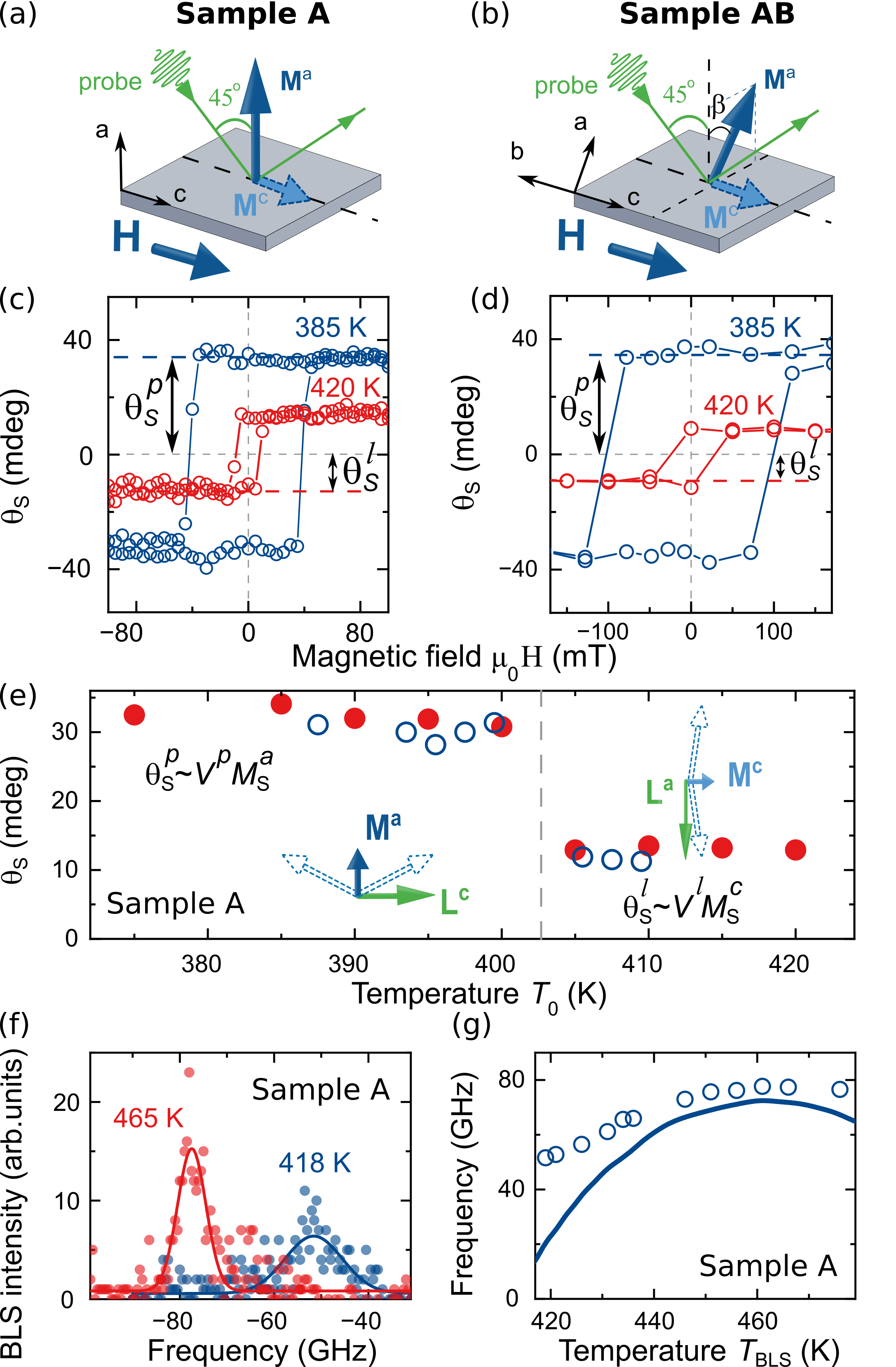}
  	\caption{\label{fig:static} 
		(a,b) Experimental geometry for magneto-optical detection of the SR transition for the (a) Sample A and (b) Sample AB. 
        Magnetic field $\mathbf{H}$ is applied close to the $c$-axis and makes a small angle (not shown) with the sample plane.
		(c,d) Equilibrium magneto-optical hysteresis loops at temperatures below ($T=385$~K, blue curves) and above ($T=420$~K, red curves) \Tsr{} for the (c) Sample A and (d) Sample AB. 
		(e) Temperature dependence of static MOKE $\uptheta^p_S$ and $\uptheta^l_S$ proportional to $M^a_S$ and $M^c_S$ below and above \Tsr, respectively for the Sample A.
        Blue and red symbols show the data obtained upon cooling down and heating up the sample, respectively.
		Switching of the magnetization by 90$^\circ$ and the change of the intersublattice angle at SR transition are schematically shown in insets.
        (f) BLS spectrum of \FBO{} at  $\upmu_{0}H$ = 200 mT and  two temperatures 418 and 465 K. Solid lines are Gaussian fits. (g) Temperature dependence of the qFMR frequency obtained from BLS (circles) and from literature data \cite{arutyunyan1989magnetic} (solid line).}
\end{figure}

For a detailed investigation of the kinetics of the SR transition, measurements were conducted on the two single-crystalline samples with different crystallographic orientations [Fig.~\ref{fig:static}(a, b)]. 
The samples were cut from bulk single crystals grown by the flux method and obtained from different sources.
In a (100)-cut sample, referred to as a Sample A, the magnetization $\mathbf{M}$ is aligned along the sample normal below the SR transition temperature \Tsr. 
Upon heating above \Tsr, $\mathbf{M}$ undergoes a reorientation into the sample plane, aligning along the in-plane $c$-axis.
In the (110)-cut Sample AB, the magnetization makes an angle $\upbeta~=~$45$^\circ$ with the sample normal below \Tsr, and is aligned within the sample plane above \Tsr. 


\subsection{\label{sec:characterization} Magneto-optical characterization}

For characterization of a magnetic state of the samples we used magneto-optical magnetometry  based on a combination of polar (pMOKE) and longitudinal (lMOKE) Kerr effects. 
Probe polarization rotation was measured with 170-fs pulses at the photon energy of 2.4 eV with the magnetic field applied close to the sample plane and to the $c$-axis, as shown in Fig.~\ref{fig:static}(a, b).
Small (about 5$^\circ$) angle the field makes with the sample plane provides a nonzero projection of the field on the sample's easy axis below \Tsr.
Probe angle of incidence of 45$^\circ$ ensures that the polarization rotation is sensitive to the in-plane magnetization above \Tsr{} through lMOKE and to the out-of-plane projection of the magnetization below \Tsr{} through pMOKE.
Probe is polarized along the $c$-axis.


Figures~\ref{fig:static}(c, d) show equilibrium MOKE hysteresis loops $\uptheta_S(\upmu_0H)$ below and above \Tsr{} measured 
in both samples. 
The loop shape at $T_0=$ 385~K indicates that the strength of the out-of-plane projection of the field is sufficient to saturate the samples along their easy axes in the \Gtwo{} phase.
In-plane field strength, in turn, is insufficient to deviate the magnetization from the $a$-axis.
Thus, at $T_0~<~$\Tsr, $\uptheta_S$ originates from the pMOKE and is proportional to the out-of-plane magnetization, $\uptheta_S^p=V^pM_S^a$ in the Sample A, and $\uptheta_S^p=\cos\upbeta\cdot V^pM_S^a$ in the Sample AB.
Above \Tsr, $\uptheta_S$ stems from the lMOKE and is proportional to the in-plane magnetization, $\uptheta_S^l=V^lM_S^c$ in both samples. 
Taking into account that $M_S^c$ = 0.6 $M_S^a$, we obtain ratios between the longitudinal $V^l$ and the polar $V^p$ magneto-optical coefficients, $[V^l/V^p]_\mathrm{A}=0.64$ for the Sample A, and $[V^l/V^p]_\mathrm{AB}=0.3$ for the Sample AB.
We note that $V^l$ and $V^p$ and their ratio in two samples are different, since they are governed by different components of the linear magneto-optical and optical susceptibilities due to different crystallographic orientations of the samples and the pronounced trichroism of \FBO{} [Sec.~\ref{sec:opt}].


Figure~\ref{fig:static}(e) shows the temperature dependence of the static MOKE signal at saturation measured for the Sample A.
The equilibrium SR transition occurs at \Tsr~=~403~K, marked by the abrupt change of $\uptheta_S$ as a result of the difference between the saturation magnetizations in the \Gtwo{} and \Gfour{} phases. 
The two curves obtained upon cooling down (blue symbols) and heating up (red symbols) coincide within experimental resolution limited by the temperature sweep step of 5~K.
This confirms that the first‑order SR transition in the sample exhibits a narrow thermal hysteresis in agreement with literature \cite{kamzin1993mossbauer,tsymbal2006orientation}.
Thus, from the equilibrium characterization it follows that, upon excitation of \FBO{} by a laser pulse at $T_0~<~T_\mathrm{SR}$, a change of MOKE magnitude at saturation along with the decrease of effective coercivity would be two signatures of the SR transition.
We note that the relatively large effective coercivity at $T~<~$\Tsr{} is a result of a small projection of the field on the corresponding easy axis.

Since the magnetization reorientation during the laser-driven SR transitions often occurs via precessional motion \cite{kimel2004laser}, the quasi-ferromagnetic resonance mode (qFMR) was measured using the Brillouin light scattering (BLS) technique in the back-scattering geometry.  
A single-mode laser with a wavelength of 532~nm was focused on the sample surface into a spot with a diameter of 25~µm using an objective lens. 
Frequency resolution was achieved using a multi-pass tandem Fabry-Perot interferometer with a Free Spectral Range (FSR) of 100 GHz and a contrast of 10$^{12}$, which allowed us to reliably resolve the inelastic magnetic scattering peaks from the strong elastic background.
The experiments were performed in a back-scattering geometry at an incidence angle of 30$^\circ$. To select the magnetic scattering cross-section and suppress the elastic background, a cross-polarization configuration was employed.
All experiments were carried out for the Sample A in the temperature range of $T_\mathrm{BLS}  = $ 418 $-$ 485 K in an external magnetic field of $\upmu_{0}H$ = 200~mT applied in the sample plane along the $c$-axis.
Figure~\ref{fig:static}(f) shows the exemplary BLS spectra obtained at two temperatures 418 and 465~K. 
By fitting these spectra with a Gaussian function to determine the central frequency, the temperature dependence of the FMR frequency was obtained [Fig.~\ref{fig:static}(g)]. 
Comparing with the literature data \cite{arutyunyan1989magnetic}, one can see qualitatively similar behavior of the temperature dependence of the frequency, although it differs quantitatively. 
This is due to the difference in the experimental methods and the presence of an external magnetic field in BLS experiments.
Below, the measured qFMR frequency is used as a reference to characterize the laser-induced magnetization dynamics as well as the phase of the material following the laser excitation.

\subsection{\label{sec:Pump-probe experiment} Pump-probe experiment}

\begin{figure}
	\includegraphics [width=0.49\textwidth] {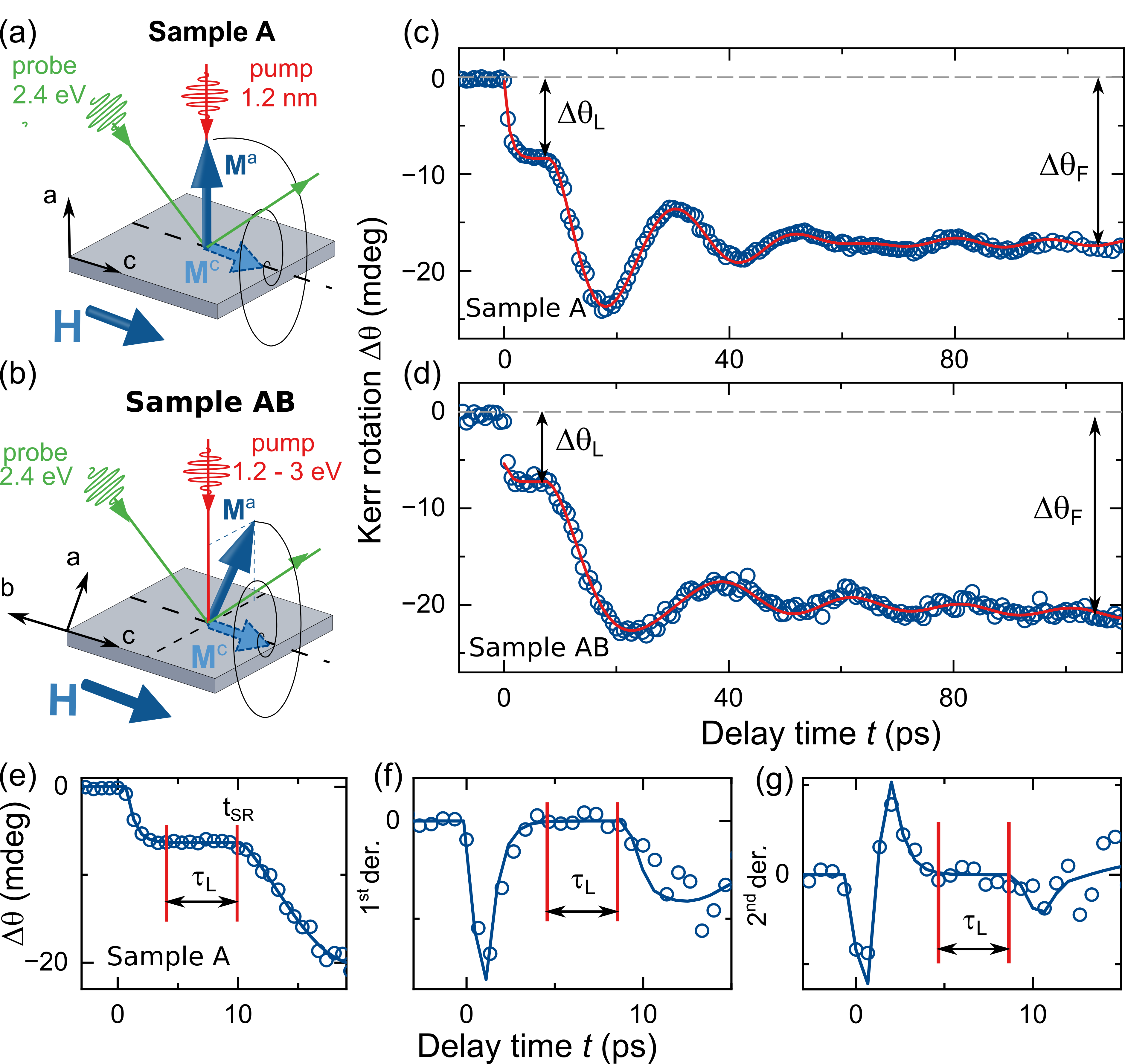}
	\caption{\label{fig:signal} 
		Experimental geometry for pump-probe experiments for (a) Sample A and (b) Sample AB. 
	Laser-induced probe polarization rotation $\Updelta\uptheta$ as a function of time delay $t$ measured in the magnetic field $\upmu_0H± = \pm$200~mT at \TO = 385~K and $F$~=~0.5~J$\cdot$cm$^{-2}$ and $E$~=~1.2~eV for (c) Sample~A and (d) Sample~AB. 
	(e) Zoomed in pump-probe signal in the time range after excitation for the Sample~A and the corresponding 
	(f) first and (d) second derivatives of the experimental signals (symbols) and of their fits (lines).
    Vertical red lines designate the time delays range where the corresponding derivative is considered to be zero.}
\end{figure}
The laser-induced SR transition in \FBO{} was investigated using a femtosecond two-color magneto-optical pump-probe setup with tunable pump photon energy. 
Laser pulses (duration: 170 fs, repetition rate: 5 kHz) were generated by a Yb$^{3+}$:KGd(WO$_4$)$_2$ regenerative amplifier PHAROS (Light Conversion) with a central photon energy $E=1.2$~eV.
The pulse was split into pump and probe pulses.
Along with the 1.2~eV excitation, additional pump pulse in the photon-energy range of $E=1.3-1.85$~eV was generated using an optical parametric amplifier (ORPHEUS, Light Conversion).
Pump pulses in this spectral range were focused at normal incidence into a spot of 60~$\mu$m diameter. 
For the pump pulses with photon energy $E=3.0$~eV, the $\beta$-barium borate (BBO) crystal was used to obtain the second harmonic from the optical parametric amplifier output at 1.5 eV. 
The 3.0~eV pump pulses were focused into a spot of 165~$\mu$m diameter.
Another BBO crystal was used to frequency-double the 1.2~eV pulses, generating probe pulses at 2.4 eV, which were the focused into a 35~$\mu$m diameter spot at an incidence angle of 45$^{\circ}$ as shown in Fig.~\ref{fig:signal}(a, b).
Incidence angle of probe pulses was kept the same as in the static measurements (Sec.~\ref{sec:characterization}) to enable comparison between the static and dynamic magnetization changes upon SR transition.
The pump fluence $F$ was varied in a range of 0.005 -- 0.6 J$\cdot$cm$^{-2}$, exceeding the probe fluence by up to a factor of 50. 

Laser-induced change of magneto-optical rotation of the probe polarization $\Updelta\uptheta$ was measured as a function of the pump-probe time delay $t$ for negative and positive sign of an external magnetic field.
To extract the field-sign dependent contribution, the signals were evaluated as
\begin{align}
\Updelta\uptheta(t) = 0.5\left[\Updelta\uptheta(H_+; t) - \Updelta\uptheta(H_-; t)\right].
\label{eq:theta}
\end{align}
Typical signals $\Delta \uptheta$ measured in the external field $\mu_0H_\pm~=~\pm200$~mT at the initial temperature \TO~=~385~K and $F$~=~0.1~J$\cdot$cm$^{-2}$ and $E~=~1.2$~eV for the Samples~A and AB are shown in Fig.~\ref{fig:signal}~(c, d).
The dynamics exhibit the same characteristic behavior in both samples, indicating the same nature of the observed phenomena. 

Transient signal $\Delta\uptheta$ for both samples demonstrates a peculiar behavior. 
A rapid decrease within $\approx$3 ps after excitation is followed by a latency period \TT{} during which the signal remains constant at a value $\Updelta\uptheta_\mathrm{L}$. 
After the latency period, at $t = t_{\mathrm{SR}}$ pronounced decaying oscillations start resulting in a signal settling at a final level $\Updelta\uptheta_\mathrm{F}$ at $t >100$~ps. 
To provide quantitative description of such behavior, we fitted $\Delta\uptheta$ transients using a piecewise function with the exponential evolution 0$\rightarrow$ $\Updelta\uptheta_\mathrm{L}$ within 0 $< t < t_{\mathrm{SR}}$, and the two damped cosine-functions superimposed by another exponential evolution $\Updelta\uptheta_\mathrm{L} \rightarrow \Updelta\uptheta_\mathrm{F}$ at $t > t_{\mathrm{SR}}$: 
\begin{equation}
    \left\{ \begin{array}{lcl}
\Updelta\uptheta_\mathrm{L}\left[ \exp\left(-\frac{t}{\uptau_{1}}\right) - 1\right]
, & \mbox{for} & t<t_\mathrm{SR}; \\

\Updelta\uptheta_\mathrm{F}\left[\exp\left(-\frac{t-t_\mathrm{SR}}{\uptau_2}\right)-1\right]
+ \\
\sum_{i=1,2}\Uptheta_{i}\exp\left(\frac{-t}{\uptau_{oi}}\right)\cos\left[2\pi f_{i}t+\upphi_{i}\right],  & \mbox{for} & t>t_\mathrm{SR}, 
\end{array}\right.\label{eq:fit}
    \end{equation}
where $\Updelta\uptheta_\mathrm{L,F}$ are the signals during the latency and at the final state; $\uptau_{1,2}$ are  characteristic times of two exponential functions.
In Fig.~\ref{fig:signal}~(e) a signal zoomed in delay time range right after excitation is shown for the Sample A.
To determine $\tau_\mathrm{L}$, the first $d (\Updelta\uptheta) /d t$ and the second $d^2 (\Updelta\uptheta) /d t^2$ derivatives of the signals and of their fits, were obtained as demonstrated in Fig.~\ref{fig:signal}~(f, g).
We extracted $\tau_\mathrm{L}$  as a time range where $d (\Updelta\uptheta) /d t$~=~$d^2 (\Updelta\uptheta) /d t^2$~=~0.
A consistent fitting procedure was applied to all experimental data, ensuring a uniform analysis across the entire set of data. 
This approach allows for a reliable comparison of the extracted parameters and confirms that the underlying physical mechanism is common to both samples.

\section{\label{sec:results} Results}

\subsection{\label{sec:results:magnetic} Magnetic field dependence}

\begin{figure}
\includegraphics [width=0.47\textwidth] {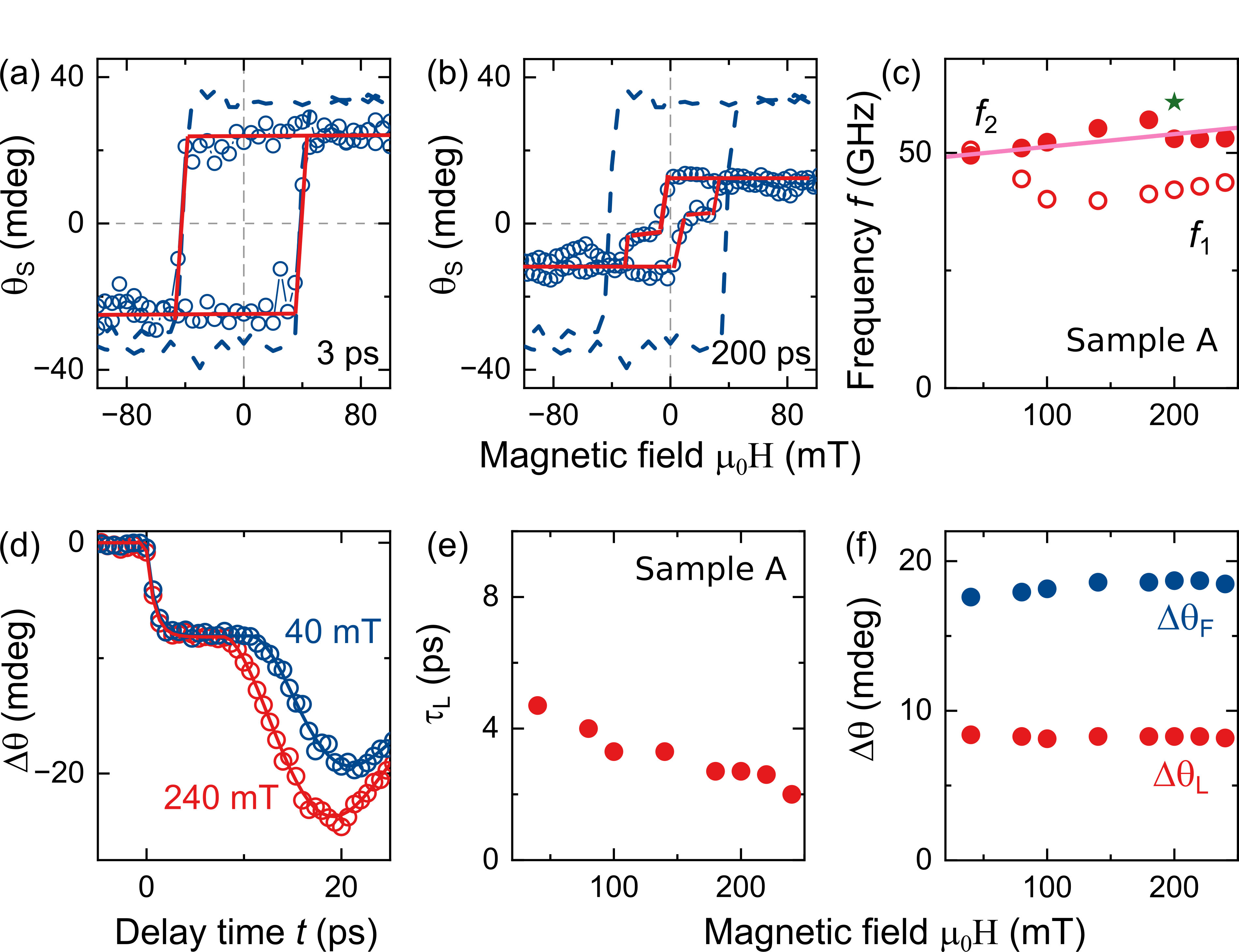}
\caption{\label{fig:magstate} 
	MOKE hysteresis loops for Sample A at the time delays (a)~$t=3$~ps and (b)~200~ps after laser excitation (symbols), and at $t=-10$~ps (dashed line). 
	Solid lines are the guides to an eye.
	(c) Field dependences of the precession frequencies $f_1$ (open symbols) and $f_2$ (closed symbols) obtained from $\Delta\uptheta(t,F)$ at \TO~=~385 K, and of the qFMR frequency (line) calculated at \TO~=~430~K for Sample A.
    Star shows the qFMR frequency at $T_\mathrm{BLS}=430$~K from BLS measurements [Fig.~\ref{fig:static}(g)].
	(d) Zoomed in transient signal during 2~ps after the excitation for the Sample A obtained at $H=40$ and 240~mT.
	Magnetic fields dependences of (e) the latency \TT, (f) $\Updelta\uptheta_\mathrm{L}$ during the latency \TT{} 
	(red symbols) and $\Updelta\uptheta_\mathrm{F}$ at $t>100$~ps (blue symbols).}
\end{figure}

We start by examining the evolution of the magnetic state underlying such magneto-optical transient at fluence $F=0.5$~J$\cdot$cm$^{-2}$ and temperature $T_0=385$~K [Fig.~\ref{fig:signal}~(c)] by analyzing how it is affected by an applied magnetic field.  
As the dynamics $\Delta\uptheta(t)$ is found to be similar in the two samples, here we here focus at the trends observed in the Sample A.
We unveil the magnetic configurations during the latency 
preceding the precession and at the final state by measuring MOKE hysteresis loops at the time delays
$t=$~3 and 200 ps [Fig.~\ref{fig:magstate}~(a, b)], respectively.  
Additionally, we verify that at negative time delay $t=-$10~ps (dashed lines in Figs.~\ref{fig:magstate}(a,b)), the hysteresis loop is the same as that measured under equilibrium conditions [Fig.~\ref{fig:static}~(c)] at the same $T_0$.
This confirms that the system relaxes back to its initial state between two subsequent pump pulses.
During the latency \TT, the coercivity is unchanged, showing that the magnetization remains aligned along the $a$-axis [Fig.~\ref{fig:magstate}(a)]. 
However, pMOKE at saturation is reduced by $\Updelta\uptheta_\mathrm{L}$.
Thus, magnetization $M^a$ is reduced at this stage by $\Delta M^a$, and $\Delta\uptheta_L=V^p\Delta M^a$. 
Fig.~\ref{fig:magstate}(d) zooms in the corresponding delay time range for two values of the magnetic field.
The latency \TT{} is found to decrease with the external magnetic field [Fig.~\ref{fig:magstate}(e)] indicating that its origin is related to the behaviour of the magnetic system after the laser excitation.
In turn, the signal magnitude during the latency $\Updelta\uptheta_\mathrm{L}$ is independent of the field as shown in Fig.~\ref{fig:magstate}(f).

At the final state, $t = $~200 ps, the signal appears to be a superposition of the two types of hysteresis loops, characteristic of the \Gtwo{} and \Gfour{} phases [Fig.~\ref{fig:magstate}(b)].
This reveals that the SR transition indeed occurs at the given pump fluence, but with only a fraction $N^\mathrm{SW}$ of the material is switched to the \Gfour{} phase.
Thus, the signal $\Updelta\uptheta_L$ at $t = $~200~ps is the sum of the pMOKE from the unswitched fraction and that of the lMOKE from the switched fraction.
As shown in Fig.~\ref{fig:magstate}~(f), $\Updelta\uptheta_L$ weakly grows with the field.

The final state is achieved at $t>100$~ps via the magnetization precession [Fig.~\ref{fig:signal}(b)].
The corresponding oscillatory signal comprises two contributions [Fig.~\ref{fig:magstate}(c)]. 
The first one possesses a higher amplitude and a lower frequency $f_1$. 
The second contribution possesses a lower amplitude and a higher frequency $f_2$. 
To compare $f_{1}$ and $f_{2}$ with the qFMR frequency in the \Gfour{} phase, we note that the external magnetic field $H$ is applied along the $c$-axis being the easy axis in this phase.
Then the frequency can be calculated using the expression $f(H)=\gamma(H + H_\mathrm{A})$ \cite{gurevich2020magnetization}, where $\gamma=$~28~GHz$\cdot$T$^{-1}$ is the gyromagnetic ratio, and $H_\mathrm{A}=2K_1(M_S^c)^{-1}$ is the effective anisotropy field assuming a contribution from biquadratic anisotropy terms entering $K_2$ being small. 
For calculations, $M^c_S=$ 1900 A$\cdot$m$^{-1}$ was taken from \cite{wolfe1969magnetization}, $K_1=$ 1.6$\cdot$ 10$^3$ J$\cdot$ m$^{-3}$ was obtained from the qFMR data at $T_0~=~$430~K in zero external field \cite{arutyunyan1989magnetic}.
The field dependence of $f_2$ [Fig.~\ref{fig:magstate}~(c)] agrees well with the calculated values of the qFMR frequency in the \Gfour{} phase at 430~K, indicating that it is a precession in the \Gfour{} phase and it possesses a linear regime.
In contrast, we associate the field-dependent frequency $f_1$ with the precession in a non-perturbative transient regime, and discuss it in details in Sec.~\ref{sec:coexistance}. 
Thus, field dependences of the transient signals reveal the three stages of the magnetic state evolution: initial change of magnetization without reorinetation, precessional switching, and quasi-steady state corresponding to coexistance of \Gtwo{} and \Gfour{} phases.

\subsection{\label{sec:Fluence} Fluence dependence}

\begin{figure}
\includegraphics [width=0.48\textwidth] {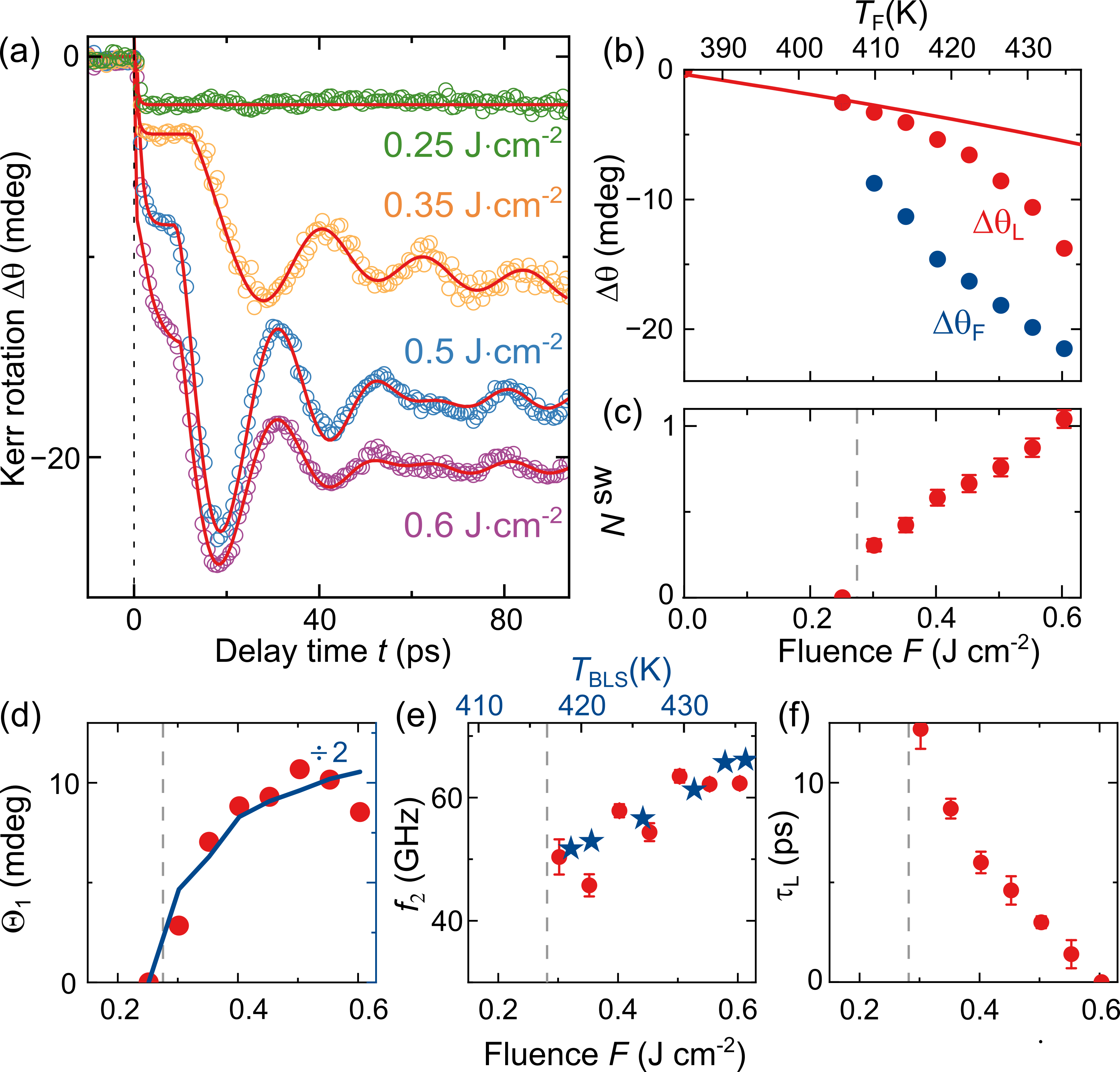}
\caption{\label{fig:Fdep}  
(a)~Pump-probe signals $\Delta\uptheta(t)$ (symbols) at $\upmu_0H~=~$200~mT, \TO~=~385~K at various pump fluences, and their fits (lines). 
(b)~Fluence dependences of $\Updelta\uptheta_\mathrm{L}$ during the latency \TT{} 
(red symbols) and  $\Updelta\uptheta_\mathrm{F}$ at $t>100$~ps (blue symbols).
Solid line is a decrease of magnetization due to laser-induced heating evaluated based on the effective temperature $T_\mathrm{F}$.
Fluence dependences of (c)~the fraction $N^\mathrm{SW}$ of the material switched to the \Gfour{} phase, (d)~the experimental and calculated amplitude $\Uptheta_\mathrm{1}$
(e)~the precession frequency $f_2$ (circles) collated with the temperature dependence of the qFMR frequency obtained from BLS (stars), 
(f)~the latency $\uptau_\mathrm{L}$.
}
\end{figure}

For laser-induced phase transitions, fluence dependences of their parameters are a key fingerprint \cite{shelukhin2022spin, afanasiev2021ultrafast}. 
Therefore, we examined $\Updelta\uptheta(t)$ traces measured for the Sample A at different $F$, fixed initial temperature $T_0=385$~K and the pump photon energy of $E=1.2$~eV [Fig.~\ref{fig:Fdep}(a)].
Evidently, $\Updelta\uptheta_\mathrm{L,F}$, \TT, and the precession parameters experience both quantitative and qualitative changes with pump pulse fluence.

As demonstrated above in Fig.~\ref{fig:magstate}~(b), at $t > 100$ ps, the magneto-optical response comprises a signal from the switched fraction of sample manifested via pMOKE, and from the non-switched fraction, manifested via lMOKE for this geometry.
It can be evaluated as
\begin{align} \label{eq:th_F}
\uptheta_F=N^{SW}  V^l  M^c+
(1-N^\mathrm{SW})\cos\upbeta\cdot V^p  (M^a+\Delta M^a),
\end{align}
where $V^l \cdot M^c=\uptheta_S^l$, and $\cos\upbeta\cdot V^p \cdot (M^a+\Delta M^a)=\uptheta_S^p+\Delta\uptheta_L$, $\upbeta=0$ for the Sample A.  
In the pump-probe experiment one measures difference between the magneto-optical signals in the laser-induced state $\uptheta_F$, and in the initial equilibrium state $\uptheta_S^p$:
\begin{align} \label{eq:final}
\Updelta \uptheta_F &=\uptheta_F - \uptheta_S^p = \uptheta_S^p - N^{SW} \uptheta_s^l - (1-N^\mathrm{SW})( \uptheta_S^p+\Delta\uptheta_L) \nonumber \\
&= N^{SW}(\uptheta_S^l - \uptheta_S^p-\Updelta \uptheta_L) +\Updelta \uptheta_L. 
\end{align}
From this relation and the experimental data of $\Updelta\uptheta_\mathrm{L}$ and $\Updelta\uptheta_\mathrm{F}$ (Fig.~\ref{fig:Fdep}(b)), we extracted the dependence of the switched fraction $N^\mathrm{SW}$ on the pump fluence $F$.
As seen in Fig.~\ref{fig:Fdep}(c) there is a threshold fluence $F_\mathrm{th}\approx 0.25$~J$\cdot$cm$^{-2}$ at which $N^\mathrm{SW}$ starts to increase from zero, that is characteristic for laser-induced phase transitions. 
Importantly, only at the highest fluence $F=0.6$~J cm$^{-2}$, the SR transition is completed, \textit{i.e.} $N^\mathrm{SW}=1$, and only the \Gfour{} phase is present.

We also examine if the precession observed at intermediate time delays is indeed related to the SR transition. According to the suggested interpretation, the amplitude of the precession $\Theta_1$ should be related to the magnetization length achieved during the latency, $M^a+\Delta M^a$ multiplied by the switched fraction $N^\mathrm{SW}$:
\begin{align} \label{eq:A1}
\Uptheta_1\propto N^{SW}  V^p  (M^a+\Delta M^a) = N^{SW} (\uptheta_S^p + \Delta \uptheta_\mathrm{L}).
\end{align}
In Fig.~\ref{fig:Fdep}(d), the experimentally obtained fluence dependence of the amplitude $\Uptheta_1$ is compared with that calculated using the relation above. 
As expected, precession occurs at $F>F_\mathrm{th}$, and the experimental and calculated fluence dependences of its amplitude exhibit qualitatively similar behavior.
Quantitative difference stems from the fact that the original fit function [Eq.~\eqref{eq:fit}] used to obtain $\Uptheta_1$ does not account for a gradual increase of the lMOKE contribution during the precession due to growing in-plane projection of magnetization.
We note that taking this process into account in a fit function would overburden the latter without yielding significant physical details.

While the fluence dependence of the precession amplitude confirms the precessional character of the switching, the precession frequency after SR transition offers a means to characterize the laser-induced phase \cite{afanasiev2021ultrafast,shelukhin2022spin,kuzikova2023laser,Kuppevelt-APL2026}.
The frequency is given by the transient magnetic parameters $M_S$, $K_{1,2}$ of the laser-induced phase, and therefore we can use the frequency $f_2$ of the precession in a linear regime to characterize this phase.
In Fig.~\ref{fig:Fdep}(e), we compare two dependences, namely the increase of $f_2$ with pump fluence and the temperature dependence of the equilibrium qFMR frequency obtained from BLS measurements. 
The comparison reveals that the effective temperature $T_\mathrm{F}=T_\mathrm{BLS}$ reached by the material after excitation for any pump fluence $F \geq F_\mathrm{th}$ exceeds $T_\mathrm{SR}$. 
On the one hand, this observation confirms the occurrence of a laser-driven SR transition. 
On the other hand, such laser-induced heating would be expected to drive a complete SR transition (i.e., $N^\mathrm{SW}=1$) already at $F_\mathrm{th}$, in accordance with equilibrium studies \cite{kamzin1993mossbauer,tsymbal2006orientation}. 
This observation reveals that laser-induced heating alone is insufficient to achieve a complete SR transition.

Apparent contradiction between the characteristics of the precession and the final state draws ones attention to the first stage of the laser-induced magnetic dynamics characterized by the parameters $\Delta\uptheta_\mathrm{L}$ and $\tau_\mathrm{L}$, which fluence dependences are presented in Figs.~\ref{fig:Fdep}(b,f).
Importantly, the latency time $\tau_\mathrm{L}$ decreases monotonically with increasing fluence and vanishes at fluences when $N^\mathrm{SW}$ reaches 1, suggesting close connection between these two parameters.
The initial magnetization decrease $\Delta\uptheta_\mathrm{L}$ observed during this stage shows a non-linear behavior.
This decrease is found to be much steeper than the one expected for a thermally-induced decrease of magnetization [see solid line in Fig.~\ref{fig:Fdep}(b)].
The latter is evaluated based on the effective temperature $T_\mathrm{F}(F)$ as 
\begin{equation}
\uptheta^p(T_\mathrm{F})-\uptheta^p_s(T_0)=\uptheta^p(0)\left( 1-\frac{T_\mathrm{F}}{T_N}\right)^\upzeta-\uptheta^p_s(T_0).\label{eq:MvsT}
\end{equation}
Here $\uptheta^p(0) \approx 54$~mdeg is pMOKE at low temperature obtained from the extrapolation of the $\uptheta^p_s(T)$ dependence [Fig.~\ref{fig:static}(e)], $\upzeta$ = 0.365 is the critical exponent of the magnetization of the 3D Heisenberg ferromagnet.

\subsection{\label{sec:Temperature} Temperature dependence}

\begin{figure}
\includegraphics [width=0.48\textwidth] {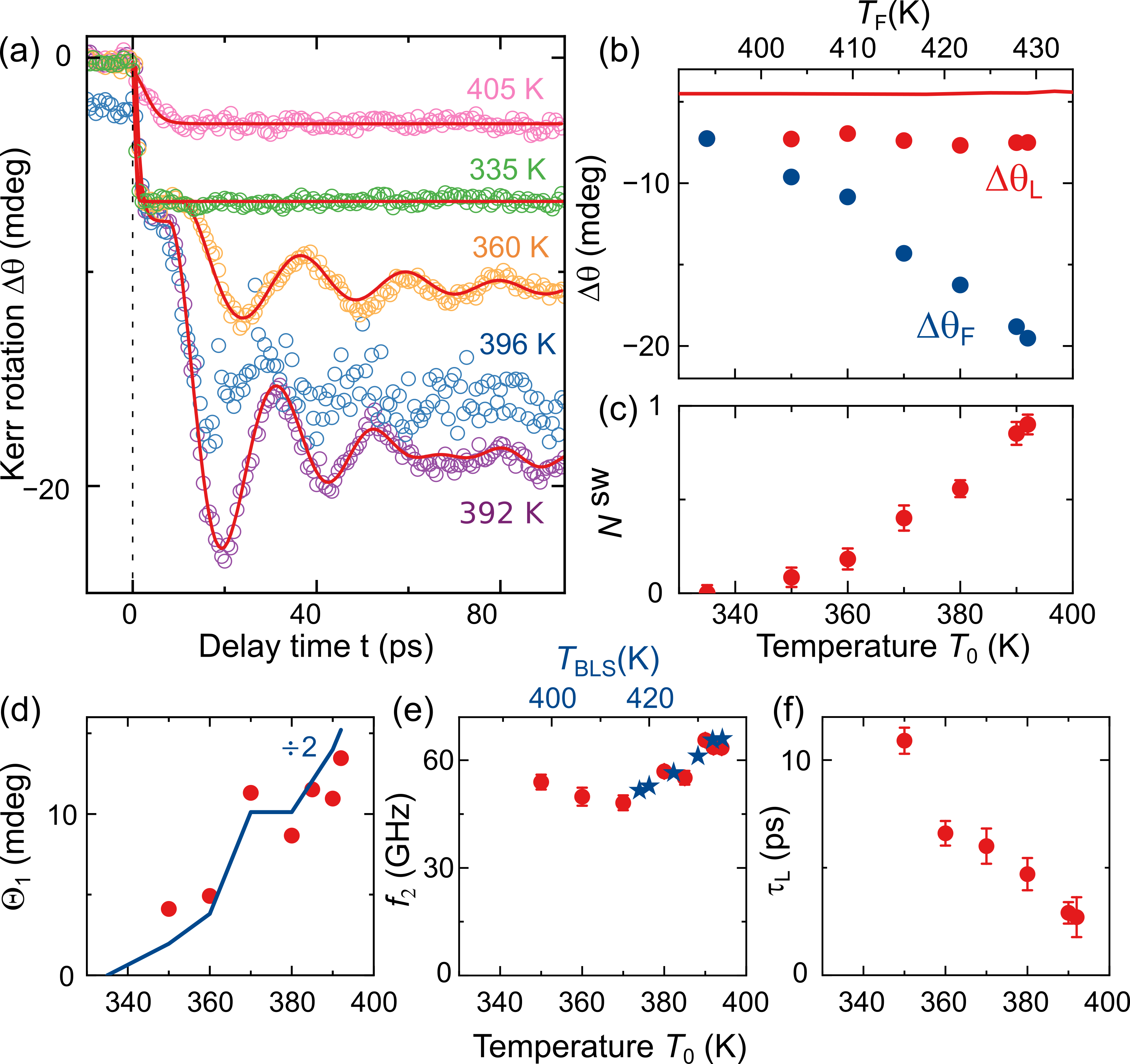}
\caption{\label{fig:Tdep} 
(a) Pump-probe signals $\Delta\uptheta(t)$ (symbols)  
measured at    ~ $F = $ 0.5 J$\cdot$cm$^{-2}$ at various initial temperatures at $\upmu_0H =$200~mT, and their fits (lines).  
(b)~Temperature dependence of $\Updelta\uptheta_\mathrm{L}$ (red symbols) during the latency \TT{} and $\Updelta\uptheta_\mathrm{F}$ at $t>100$~ps (blue symbols).
Solid line is a decrease of magnetization due to laser-induced heating evaluated based on the effective temperature $T_\mathrm{F}$ [Eq.~\eqref{eq:MvsT}].
Temperature dependence of (c)~the switched fraction $N^\mathrm{SW}$, (d)~the experimental (symbols) and calculated (line) amplitude $\Uptheta_\mathrm{1}$, (e)~the precession frequency $f_2$ (circles) collated with the temperature dependence of the qFMR frequency obtained from BLS (stars),  and (f)~the latency $\uptau_\mathrm{L}$.}
\end{figure}

A combined analysis of the evolution of the signals and their parameters with initial sample temperature is performed to further confirm that the SR transition is observed and to evaluate the role of laser-induced heating.
Figure~\ref{fig:Tdep}(a) shows the pump-probe signals measured at various initial temperatures at fixed fluence $F$ = 0.5 J$\cdot$cm$^{-2}$ for Sample A. 
All signatures of the laser-induced SR transition are observed only in the range of 335~K$~< T_0 < T_\mathrm{SR}$,  when the sample is in the \Gtwo~ phase initially and the fluence is sufficient to induce the SR transition. 
At lower or higher fluence the lowest limit shifts to higher or lower temperature, respectively, as was verified at the fluence value $F=0.35$J$\cdot$cm$^{-2}$.

The temperature dependences of the parameters $\Updelta\uptheta_\mathrm{L}$ and $\Updelta\uptheta_\mathrm{F}$, describing the magnetic state during initial stage  $\uptau_\mathrm{L}$ and at the final time $t~>~$ 100 ps, respectively, are plotted in Fig.~\ref{fig:Tdep}(b).
Using these values and Eq.~\eqref{eq:final}, we calculated the switched fraction $N^\mathrm{SW}$ [Fig.~\ref{fig:Tdep}(c)].
Next, using  Eq.~\eqref{eq:A1}, values $N^\mathrm{SW}$, $\uptau_\mathrm{L}$ and $\uptheta^p_s$  we calculated the amplitude $\Uptheta_1$ [Fig.~\ref{fig:Tdep}(d)].
The precession frequency during the SR transition is used to estimate the final temperature $T_\mathrm{F}$ by collating two dependences, namely the laser-induced precession frequency $f_2$ and the equilibrium qFMR frequency obtained from BLS [Fig.~\ref{fig:Tdep}(e)].
As in the case of the fluence dependence, laser-induced heating above \Tsr{} appears to be insufficient for reaching the complete switching, $N^\mathrm{SW}=1$. 
The latency $\uptau_\mathrm{L}$ [Fig.~\ref{fig:Tdep}(f)] shows a decrease with temperature, while $\Updelta\uptheta_\mathrm{L}$ grows with the temperature only weakly, and its magnitude exceeds the one estimated using Eq.~\eqref{eq:MvsT} from a thermally-induced demagnetization [see solid line in Fig.~\ref{fig:Tdep}(b)].

At $T_0 >$ \Tsr~ the transient $\Updelta\uptheta(t)$ does not show any signatures of the laser-driven SR transition at any studied fluence, as shown in Fig. \ref{fig:Tdep}(a) for $T_0 =405$~K.
The transient signal consists of initial decrease $\Updelta\uptheta$ and only its magnitude grows with fluence. 

When the initial temperature $T_0$ is approaching but not exceeding $T_\mathrm{SR}$, two additional effects emerge, both characteristic of the laser-driven SR transitions [see data for $T_0~=~$396~K in Fig.~\ref{fig:Tdep}(a)]. 
First, at negative time delays, the signal $\Updelta\uptheta$ is not equal to zero.
At positive delays, the transient signal at this temperature demonstrates additional fluctuations.
Note that at $T_0 >$ \Tsr~ both additional effects disappear.
We ascribe such behavior to the fact that, at $T_0$ being in the vicinity of $T_\mathrm{SR}$, the excited system returns to different initial states after each pump pulse.
Then the system is excited to a distinct state by a subsequent pulse.
This is due to the fact that the first-order phase transitions are characterized by a sharp change in the thermodynamic properties of the system when passing through the transition temperature, when the system stability is lost and fluctuations are greatly increased \cite{weiss2023discovery}.
Thus, the closer the initial sample temperature to the transition, the larger is the contribution to the transition from increasing fluctuations.    

\subsection{\label{sec:Pump_energy} Pump photon energy dependence}

\begin{figure}[h]
	\includegraphics [width=0.48\textwidth] {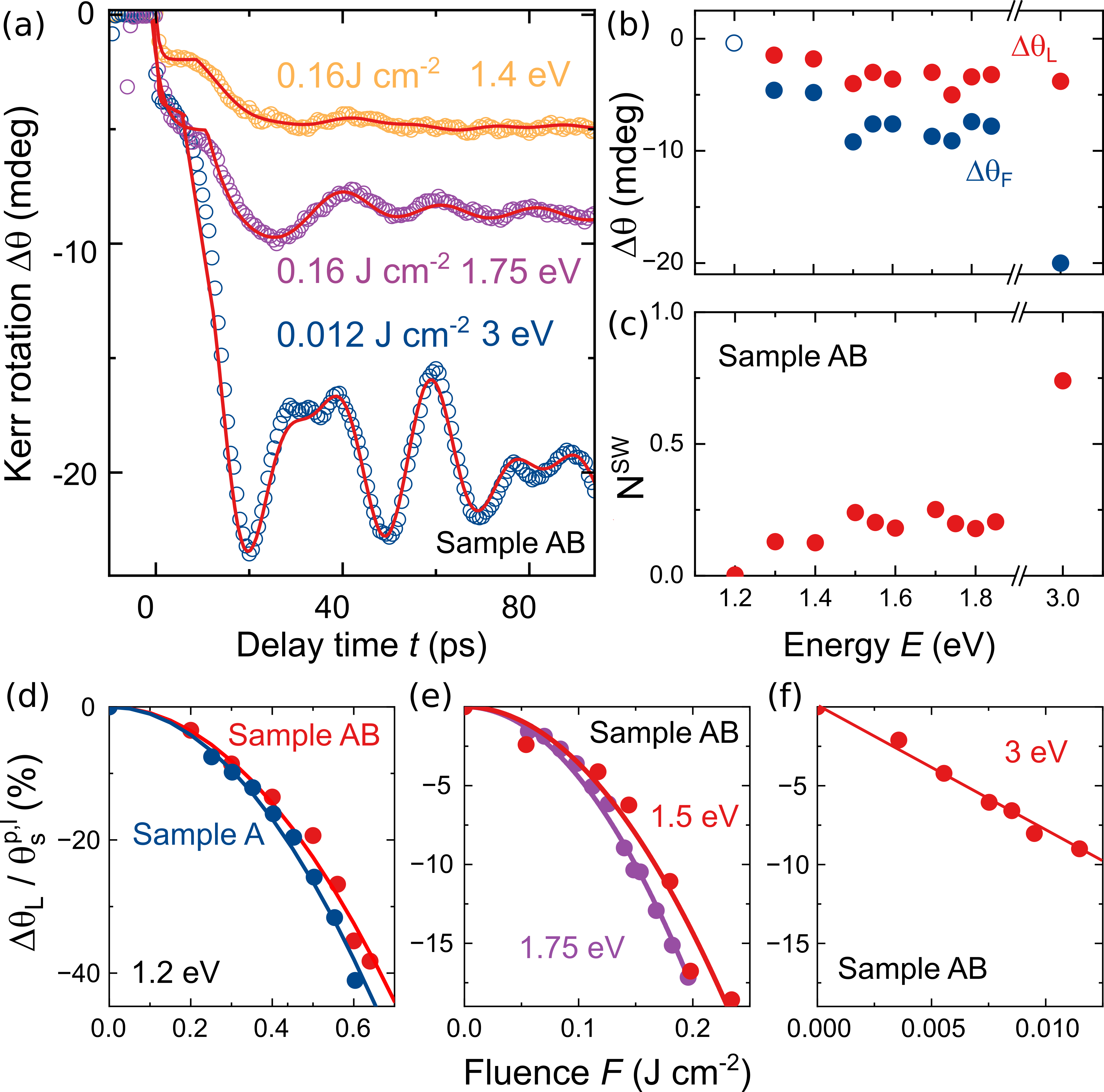}
	\caption{\label{fig:Edep}
		(a) Pump-probe signals $\Delta\uptheta(t)$ (symbols)  
		measured at  initial temperature \TO=385~K and at $\upmu_0H =$200~mT for pump photon energy $E=1.3-1.85$~eV at ~ $F = $ 0.16 J$\cdot$cm$^{-2}$ and for pump photon energy $E=3.0$~eV at $F = $ 0.012 J$\cdot$cm$^{-2}$. 
		Pump photon energy dependences of (b)~$\Updelta\uptheta_\mathrm{L}$ (red symbols) and $\Updelta\uptheta_\mathrm{F}$ (blue symbols), and (c) $N^\mathrm{SW}$.
        Pump fluence was of $F=0.012$~J$\cdot$cm$^{-2}$ at $E=3.0$~eV and $F=0.16$~J$\cdot$cm$^{-2}$ at $E=1.2,~1.3-1.85$~eV.
        Experimental dependence (symbols) of $\Updelta\uptheta_\mathrm{L}/\Uptheta^\mathrm{p}_\mathrm{S}$ on pump fluence for Sample AB at \TO=385~K for pump photon energies (d) $E=1.2$~eV, (e) 1.5 and 1.75~eV, and (f) 3.0~eV, and their fits (lines) using functions $-F^2$ (d,e) and $-F$ (f).
        In (d) the same dependence is shown for the Sample A as well.}
\end{figure}

Further insight into the laser-induced transition can be gained based on the changes of transient signals with the pump pulse photon energy. 
Three spectral ranges are chosen based on optical absorption spectrum [Fig.~\ref{fig:structure}(b)].
At $E~=~1.2$~eV, there is no resonant absorption, and the absorption coefficient $\upalpha=280$~cm$^{-1}$ is the lowest.
In the range 1.3-1.85~eV, we scan through the $d-d$ transition, with $\upalpha$ taking moderate values between 450 and 1050~cm$^{-1}$.
At 3.0~eV, there is a fundamental absorption due to charge-transfer (CT) transitions 2$p$(O$^{2-}$)$\rightarrow$3$d$(Fe$^{3+}$), with $\upalpha=1.87\times10^{5}$~cm$^{-1}$.
Figure~\ref{fig:Edep}(a) shows the $\Updelta\uptheta(t)$ traces measured at different pump photon energy $E$ at the fixed initial temperature $T_0~=~$385~K for the Sample~B.
For this plot, pump fluences were chosen such that $\Updelta\uptheta_{L}$ has the same value of -3~mdeg at the high-energy end of intermediate spectral range ($E~=~1.85$~eV) and at $E=3.0$~eV. 
For $E~=~1.3-1.85$~eV and 3.0~eV the fluence was of $F~=~$~0.160~J$\cdot$cm$^{-2}$ and 0.012~J$\cdot$cm$^{-2}$, respectively.

Figures~\ref{fig:Edep}(b-c) show the parameters $\Updelta\uptheta_{L,F}$, as well as $N^\mathrm{SW}$ as functions of the pump photon energy $E$.
All three parameters exhibit similar dependences in the ranges of low and moderate absorption, with weak growth as $E$ increases and nearly constant values in the range $1.5-1.85$~eV.
Noticeably, at $E=3.0$~eV this correlation between $\Updelta\uptheta_{L}$, $\Updelta\uptheta_{F}$, and $N^\mathrm{SW}$ is broken down, with $\Updelta\uptheta_{F}$ and $N^\mathrm{SW}$ reaching much larger values at the same value of $\Updelta\uptheta_{L}$.
This result further substantiates the assumption that there is no simple connection between the laser-induced heating related to absorption $\upalpha$ and the behavior of $\Updelta\uptheta_{L}$, $\Updelta\uptheta_{F}$, and $N^\mathrm{SW}$.
Finally, fluence dependences $\Updelta\uptheta_{L}$ observed at different pump photon energies and shown in Fig.~\ref{fig:Edep}(d-f) appear to have different character depending whether the pump pulse is in the range of the strong fundamental absorption or not.
Only in the former case [Fig.~\ref{fig:Edep}(f)] this dependence is linear, while in other cases it follows $\propto-F^2$ [Fig.~\ref{fig:Edep}(d-e)].

Overall, experimental results presented in this Section~\ref{sec:results} provide clear evidence of a nontrivial laser-induced SR transition in \FBO. 
The transition occurs via precession and is only partial in a certain range of fluences and initial sample temperatures, 
so that a fraction of the material $N^\mathrm{SW}$ transforms into the phase \Gfour{} while the rest remains in the \Gtwo{} phase.
But the fluence and temperature dependences of $N^\mathrm{SW}$ cannot be explained based on a laser-induced heating only.
The switching is preceded by a net magnetic moment decrease $\Delta M^a\propto \Updelta\uptheta_\mathrm{L}$ that proceeds without rotation and persists during a time interval \TT.
$\Delta M^a$ does not correlate with laser-induced heating, and shows different fluence dependences depending on the pump photon energy. 
The latency time \TT{} itself decreases with fluence, temperature, and the applied magnetic field.
All this observations suggest that the first stage of the process represents a bottle-neck for the complete transition.

\section{\label{sec:discution} DISCUSSION}

\subsection{Laser-induced change of the parameters responsible for the spin canting}

We first discuss the physical origin of the first stage of the magnetization evolution. 
As discussed above, during the time interval \TT{} the signal $\Updelta\uptheta_\mathrm{L}$ is proportional to a decrease in the magnetization $\Updelta M^a$, without its rotation by $90^\circ$. 
This decrease cannot be explained by laser-induced heating leading to demagnetization, as demonstrated in Fig.~\ref{fig:Fdep}(b) and Fig.~\ref{fig:Tdep}(b), where the calculated decrease of magnetization (solid lines) does not reproduce the experimental data. 
We therefore suggest a mechanism for the observed reduction of the net magnetic moment due to a change in the intersublattice canting. 

\begin{figure}[t]
	\includegraphics [width=0.48\textwidth] {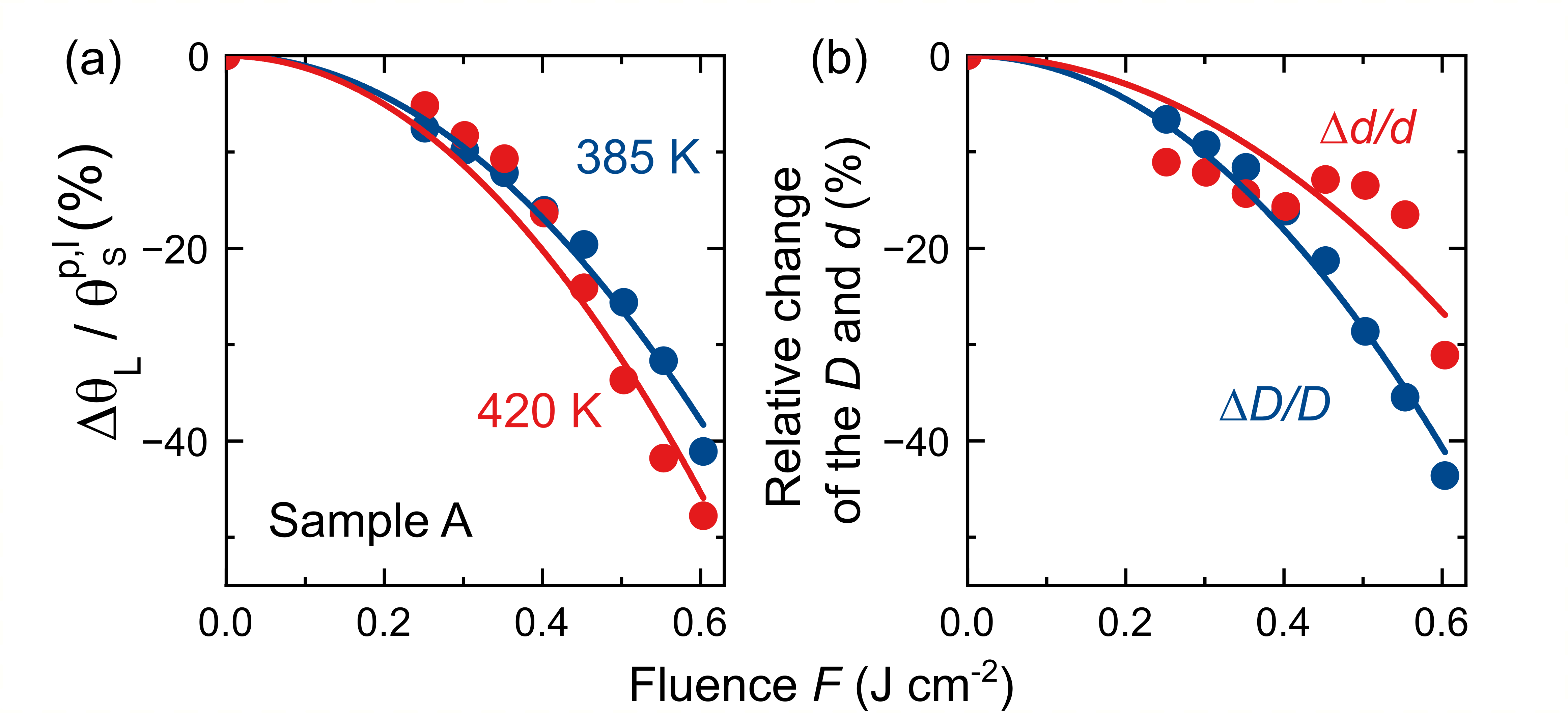}
	\caption{\label{fig:DMI}
			(a) The ratio of the change $\Updelta\uptheta_\mathrm{L}$  to the  $\uptheta^{p,l}_S$, plotted as a function of pump fluence for the Sample A, measured at $\upmu_0H =$~200~mT and $E=1.2$~eV for \TO~=~385~K (blue symbols)  and \TO~=~420~K (red symbols) 
		(b) Relative change of the parameters $D$ and $d$ as a function of pump fluence as evaluated based on $\Updelta\uptheta_L(F)$ using Eq.~\eqref{eq:dD}.
        Lines are the fits using $\propto -F^2$ function.}
\end{figure}

The decrease of the canting between the sublattices can result from either decrease of the two- and single-site contributions to the weak ferromagnetism, $D$ and $d$, or from the increase of the exchange parameter $E$ [see Eq.~\eqref{eq:M}].
Distinguishing between these two processes in a weak ferromagnet is challenging \cite{mikhaylovskiy2015ultrafast}.
However, in the present case one could argue that the laser-induced change of exchange coupling is not feasible.
Indeed, the laser-induced change $\Delta M^a/M^a_S=\Delta\uptheta_L/\uptheta^p_S$ achieved at large fluences is of 50~\% [Figs.~\ref{fig:Edep}(d-f),\ref{fig:DMI}(a)], requiring a corresponding increase of the intersublattice exchange coupling by nearly a factor of two.
Thus, we consider the change of $D$ and $d$ as the most feasible underlying process. 

Experimental dependences $\Updelta\uptheta_L(F)$ below and above \Tsr{}, and the ratio of the equilibrium parameters $d\approx0.25D$ (see Sec.~\ref{sec:SR}), enable us to evaluate the relative changes of both parameters $D$ and $d$.
Indeed, from Eqs.~\eqref{eq:M} we obtain 
\begin{align}
    \frac{\Delta\uptheta_L}{\uptheta^{p(l)}_S} = \frac{\Delta D \pm \Delta d}{D \pm d} \label{eq:dD}
\end{align}
where $\pm$ and superscript $p(l)$ correspond to the case below (above) \Tsr. 
Figure~\ref{fig:DMI}(a,b) shows corresponding fluence dependences obtained at $T_0=385$ and 420~K for the Sample A and with the pump at the photon energy $E=1.2$~eV.
Both parameters $D$ and $d$ decrease with fluence with their relative changes reaching $\sim40$~\% and  $\sim30$~\%, respectively.
However, the equilibrium value of $D$ is approximately four times larger than that of $d$ [Sec.~\ref{sec:SR}].
Thus, despite the comparable relative changes of the two parameters, the laser-induced variation of $D$ provides the leading contribution to the observed change of the angle between antiferromagnetic sublattices.

Both, $\Updelta D$ and $\Updelta d$ exhibit quadratic dependence on the laser fluence for photons at $E=1.2$~eV [Fig.~\ref{fig:DMI}(b)].
Analogous conclusion can be drawn from the data obtained at the pump photon energies in the range of intermediate absorption [Fig.~\ref{fig:Edep}(e)].
At $E=3.0$~eV, $\Updelta D,~\Updelta d$ exhibit linear dependences based on the corresponding dependence of $\Updelta\uptheta_L(F)$ [Fig.~\ref{fig:Edep}(e)].
This unambiguously reveals that the CT transition from $2p$ states of O$^{2-}$ to $3d$ states of Fe$^{3+}$, namely $t_{2u}(\pi)\rightarrow t_{2}$ centered at ~3.0~eV \cite{pisarev2009charge}, is required to enable the change of these parameters. 
Indeed, at $E=1.2;~1.3-1.85$~eV such transition requires a two-photon absorption [Fig.~\ref{fig:static}(b)], which is in a full agreement with the fluence dependence $\propto F^2$.
At $E=3.0$~eV the pump pulses are resonant with the CT transitions  [Fig.~\ref{fig:structure}(b)], and the one-photon absorption takes place.
Also, pump fluences required to reach the same values of $\Updelta D,\Updelta d$ and $\Updelta M^a\propto\Updelta\uptheta_L$ are an order of magnitude lower for the direct absorption at $E=3.0$~eV as compared to the two-photon absorption.

We observe no distinct features in the spectral dependence of $\Updelta\uptheta_L$ [Fig.~\ref{fig:Edep}(b)] that could be attributed to electronic $3d-3d$ transition centered at around 1.5~eV.
Their possible contribution to the magnetic response appears to be strongly suppressed compared to the dominant two-photon absorption process, and therefore does not produce a discernible feature in the measured signals.
One the one hand, $d-d$ transitions change the electrons distribution within $3d$ orbitals of Fe$^{3+}$ affecting its effective orbital momentum, orbitals overlap with O$^{2-}$.
This can change both the DMI parameter $D$, as well as the single-ion parameter $d$.
On the other hand, CT transitions also change the number of electrons on $3d$ orbitals of Fe$^{3+}$ thus impacting DMI and single-ion anisotropy though altering overlap with oxygen orbitals, effective orbital momentum, and magnetic moment.
Our results of spectral and fluence dependences of $\Updelta\uptheta_\mathrm{L}$ clearly distinguishes between the two pathways to alter $D$ and $d$ and shows the leading role of the CT transition at least in the regime of strong excitation.

It is worth noting that the fluences used in our experiments in the low and moderate absorption are of order of fractions of J$\cdot$cm$^{-2}$, that is considerably higher than the typical fluences used in most of experiments on laser-driven change of spin canting.
Using such fluences is enabled by a high initial temperature in the experiments leading to a large heat capacity and, as a result, relatively low laser-induced heating.
We argue that high fluences facilitated reaching large values of laser-induced modulation of $D$ and $d$.
It also allowed dominating contribution from the two-photon absorption \cite{zeuschner2020measurement, chaban2022nonlinear} and CT-based mechanism across the whole studied range of the pump pulse photon energies

The evolution of the spin canting triggered by the change of $D,d$ unfolds on a picosecond timescale. 
$\Updelta\uptheta$ reaches $\Updelta\uptheta_\mathrm{L}$ within 3-6~ps depending on the pump fluence and initial temperature [Fig.~\ref{fig:Fdep}(a) and~\ref{fig:Tdep}(a)]. 
This timescale agrees with the softening of the quasi-antiferromagnetic resonance (qAFMR) mode, which decreases from $\approx$~300~GHz at \TO~=~385~K \cite{arutyunyan1989magnetic}. 
Given that such a softening can be induced by a reduction of $d_a=D+d$, we suggest that the spin canting evolves via a single cycle of strongly damped qAFMR \cite{mikhaylovskiy2015ultrafast}. 
This implies that the characteristic time for the decrease of $D,d$ is comparable to or shorter than 3--6~ps.

$\Updelta\uptheta$ transients obtained at $F~<~F_\mathrm{th}$ at $T_0~<~ $\Tsr{} [Fig.~\ref{fig:Fdep}(a)] and at all fluences used in our study at $T_0~>~$\Tsr{} [Fig.~\ref{fig:Tdep}(a)] suggests that the laser-induced change of $D,~d$ persists on a time scale of beyond 100~ps.
Such long timescale is justified by the fact that the CT transitions are responsible and the corresponding excited state may have a long relaxation times.
Notably, such a long timescale of $\Updelta D,~\Updelta d$ is in agreement with recent results on transient change of the DMI parameter in a weak ferromagnet ErFeO$_3$ \cite{Kuppevelt-APL2026}, obtained by studying qAFMR frequency under excitations by laser pulses with $E=1.55$~eV and a fluence below that used in our experiments.

\subsection{\label{sec:latency}Phase boundaries for the laser-induced SR transition}

Associating the initial stage of the magnetization evolution to the laser-induced spin canting enables us to comprehend the apparent bottle-neck for the transient SR transition in \FBO{}, i.e. laser-induced heating being insufficient for complete transition.
We recall that at equilibrium, SR transition involves a change of the weak ferromagnetic moment due to a change of the canting of the sublattice magnetizations [Eq.~\eqref{eq:M}].
Based on the phenomenological approach outlined in Sec.~\ref{sec:SR}, we now define the SR boundaries [Eq.~\eqref{eq:bound}] in the case of the laser-induced transition.
As established in our experiments, right after the excitation the magnetization remains aligned along the $a$-axis.
Thus, during the latency, the energy profile is determined by the magnitude of $M^{a}=M_0m^{a}$.
As changes of $M_0$ are supposed to be small, we focus on the effect of $m^a$.
To trigger the precession, the minimum corresponding to the \Gfour{} should emerge [Fig.~\ref{fig:structure}(d)].  
In particular, such local minimum require $K_1(m^a)\geq0$, that is analogous to the condition at equilibrium.
Once this condition is satisfied via laser-induced change of parameters entering $K_1$, partial switching to \Gfour{} phase is possible via precession, and $0\leq N^\mathrm{SW}\leq1$.
However, the complete switching requires $K_1(m^a)+2K_2(m^a)\geq0$, that is different from the equilibrium condition. 
As a result, the effective temperature $T_F$ reached after the excitation does not serve as a proper merit for the laser-induced SR transition, as evident from the dependence of $N^\mathrm{SW}(F)$ [Fig.~\ref{fig:Fdep}(c)] and from the estimated heating $T_F$[Fig.~\ref{fig:Fdep}(e)].
Condition $K_1(m^a)+2K_2(m^a)\geq0$ is now controlled by the spin canting achieved via the change of $D, d$.  

\begin{figure}[h]
	\includegraphics [width=0.48\textwidth] {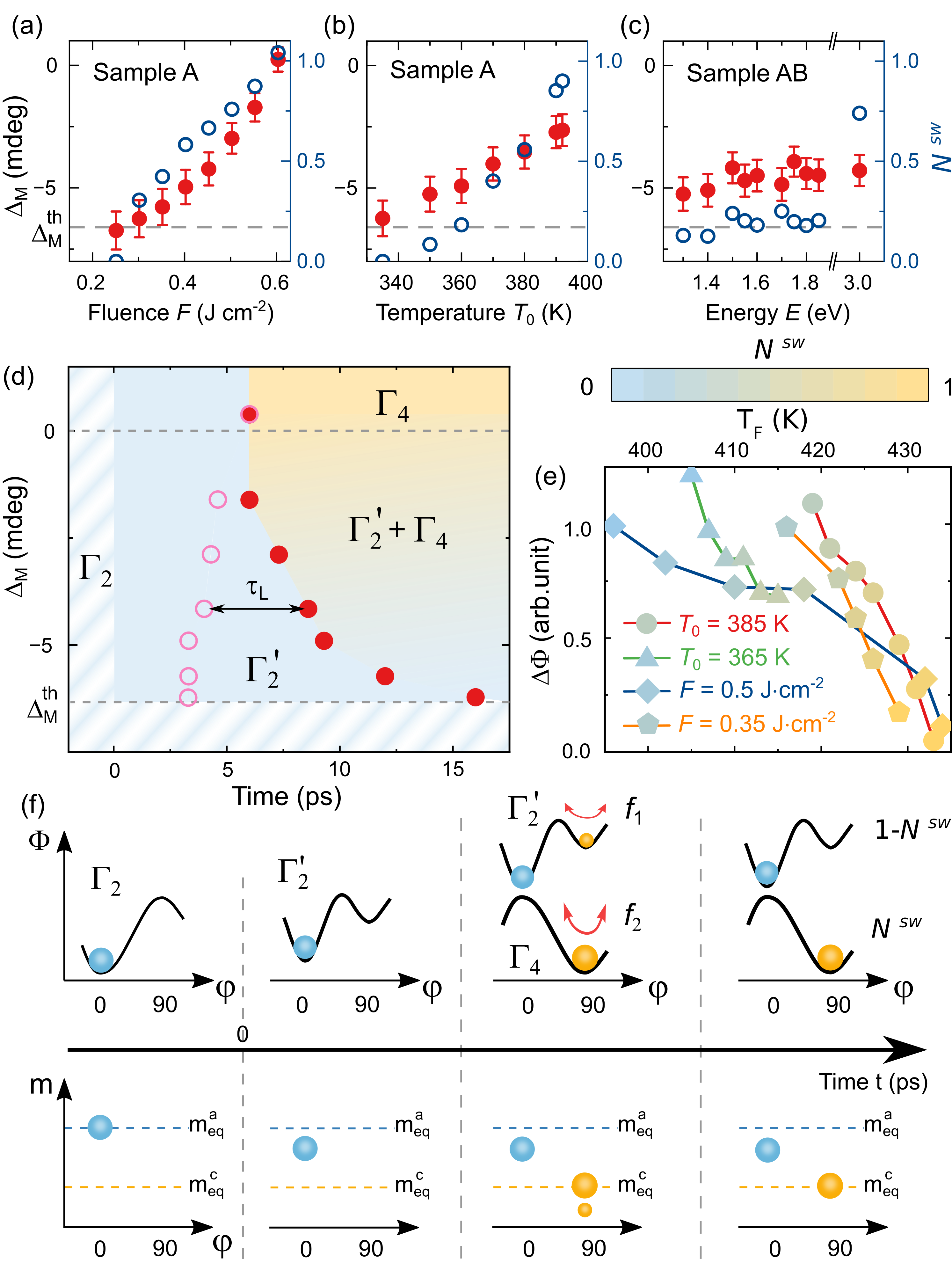}
	\caption{\label{fig:phase}
		(a)~Fluence, (b)~temperature, and (c)~pump photon energy dependences of $\Updelta_\mathrm{M}$ during the latency \TT{} and of the switched fraction $N^\mathrm{SW}$.
        Data in panels (a) and (b) are obtained at $E=1.2$~eV at (a) $T_0$ = 385~K and (b) $F=0.5$~mJ$\cdot$cm$^{-2}$.
        Data in panel (c) are obtained at $T_0$=385~K, for the same fluences as in Fig.~\ref{fig:Edep}(b).
		(d)~Phase diagram~$t-\Delta_\mathrm{M}$ of the laser-induced SR transition based on the fluence dependence at $T_0=385$~K with the color-coded switched fraction $N^\mathrm{SW}$.
		Open and closed symbols mark the start and the end (\tsr) of the latency.
       (e)~2D plot of $N^\mathrm{SW}$ vs. barrier $\Delta\Phi$ and the final temperature $T_\mathrm{F}$ as obtained from the fluence dependences at  $T_0$=385 (circles) and 365 K (triangles) and the temperature dependeces at $F$ = 0.5 (diamonds) and 0.35 ~J$\cdot$cm$^{-1}$(pentagons). 
       (f) Schematic illustration of evolution of the energy profile $\Upphi(\upphi)$ following laser excitation and the corresponding values of magnetization $m^{a(c)}$.}
\end{figure}

To relate the boundary conditions to the experimental values $\Updelta\uptheta_L\propto\Updelta M^a$, we introduce a magnetization mismatch parameter $\Updelta_\mathrm{M}$ characterizing a difference between the magnetization $M_S^c$ in the equilibrium \Gfour{} phase and $[M^a_S+\Updelta M^a(F)]$ at $t<~$\tsr.
This difference is multiplied by $V^l$ that enables evaluation of $\Updelta_\mathrm{M}$ based on the measured equilibrium and transient MOKE values. 
Thus, we obtain:
\begin{align}
    \Delta_\mathrm{M}=\uptheta_S^l-\frac{V^l}{V^p \cdot \cos\upbeta}\left[\uptheta_S^p+\Delta\uptheta_\mathrm{L}\right].\label{eq:mismatch}
\end{align}
This fluence dependence is plotted in Fig.~\ref{fig:phase}(a) for the Sample A and the pump photon energy $E=1.2$~eV.
We chose this set of data as the one with minimal laser-induced heating due to the lowest absorption [Fig.~\ref{fig:structure}(b)].

The magnetization mismatch $\Delta_\mathrm{M}^\mathrm{th}$ at $F_\mathrm{th}$ is found to be equal to the discontinuity of the equilibrium magneto-optical signal at \Tsr{} [Fig.~\ref{fig:static}(e)] expressed according to Eq.~\eqref{eq:mismatch}.
The complete laser-induced SR transition ($N^\mathrm{SW}=1$) is realized when $\Updelta_\mathrm{M}=0$, i.e. 
$[M^a_S+\Updelta M^a(F)]$ matches $M_S^c$.
Concomitantly, the latency \TT{} reduces to zero [Fig.~\ref{fig:Fdep}(e)].
This fully agrees with the defined above boundaries for the laser-induced transition.

\subsection{Switching scenario and the role of laser-induced heating}

The analysis demonstrates that the change of 
spin canting by a laser pulse with the fluence above $F_\mathrm{th}$ prepares the state with the two-minima potential, corresponding to $K_1>0$ and $K_1+2K_2<0$ [Fig.~\ref{fig:structure}(d)]. 
For the switching of the magnetization from the $a$-axis ($\upvarphi=0$) to the $c$-axis ($\upvarphi=90^\circ$), the barrier between the minima should be overcome.
The height of the barrier $\Delta\Phi$ derived from Eq.~\eqref{eq:energy} is:
\begin{align}
    \Delta\Upphi=-\frac{K_1^2}{4K_2}-K_1-K_2-\mu_0M_0mH\sqrt{1+\frac{K_1}{2K_2}}\label{eq:barrier}
\end{align}
and is controlled mainly by $\Updelta_\mathrm{M}\propto M^a$ via $K_{1}$ and by an applied magnetic field.
The barrier height spans from $-K_2-\mu_0M_0H$ to 0 as $K_1$ takes a values corresponding to the two phase boundaries.
Following the N\'eel-Brown model for a macrospin switching \cite{brown1963thermal}, the ratio between $\Delta\Phi$ and the thermal energy $k_\mathrm{B}T_\mathrm{F}$ defines the probability $N^\mathrm{SW}=1-\exp(-t/\tau_L)$ and a characteristic time required for switching between the two minima.
Here we take the latency \TT{} as a characteristic time and express it as
\begin{align}
	\uptau_{\mathrm{L}} \propto \mathrm{exp} \left[ \frac{\Updelta\Upphi}{k_{\mathrm{B}} T_{\mathrm{F}}}\right].\label{eq:tauL}
\end{align}

In particular, the growth rate of $N^\mathrm{SW}$ with increase of $\Delta_\mathrm{M}$ should change when the range of the corresponding $T_\mathrm{F}$ is changed.
This is realized in an experiment performed at various initial temperatures $T_0$ and fixed fluence $F=0.5$~J$\cdot$cm$^{-1}$ for the Sample A [Fig.~\ref{fig:phase}(b)].
Since $\Delta\uptheta_L$ is of non-thermal origin, its value is defined by the fluence and only weakly depends on $T_0$ [Fig.~\ref{fig:Tdep}(b)]. 
Correspondingly, the magnetization mismatch $\Delta_\mathrm{M}$ is weakly altered upon increase of $T_0$ [Fig.~\ref{fig:phase}(b)], while $T_\mathrm{F}$ shows a pronounced linear growth with $T_0$.
As a result $N^\mathrm{SW}$ shows a pronounced increase with $T_0$ and complete switching is realized at nonzero magnetization mismatch [Fig.~\ref{fig:phase}(b)].
This thermally activated switching is further confirmed by the data obtained at different pump photon energies $E=3.0$~eV.
Figure~\ref{fig:phase}(c) sums up the values of $\Updelta_\mathrm{M}$ and $N^\mathrm{SW}$ obtained at \TO=385~K at different pump photon energies with fluences chosen in a way that the values of $\Updelta_\mathrm{M}$ are the same at $E=3.0$~eV and 1.85~eV.
Evidently, at $E=3.0$~eV, strong absorption and higher $T_F$ lead to larger $N^\mathrm{SW}$ at the same $\Updelta_\mathrm{M}$, as compared to the results obtained at medium or low absorption. 

To show the agreement between the temperature and fluence dependences, we use Eq.~\eqref{eq:tauL} that accounts for the variations of \TT{} with fluence and temperature. 
The sets of \TT{} and $T_\mathrm{F}$, along with $N^\mathrm{SW}$, were obtained from the fluence dependences measured at \TO~=~365 and 385~K and from the temperature dependences at $F$~=~0.35 and 0.5~J$\cdot$cm$^{-2}$ for the Sample A. 
Then $\Delta\Upphi$ was evaluated based on Eq.~\eqref{eq:tauL}.
In Fig.~\ref{fig:phase}(e) we plot resulting dependence of $N^\mathrm{SW}$ on $\Upphi$ and $T_\mathrm{F}$ showing agreement between different data sets.

Suggested switching scenario is also supported by the decrease of \TT{} with an applied magnetic field [Fig.~\ref{fig:Fdep}(e)].
Indeed, as the field is applied along the $c$-axis, it reduces the barrier between the two minima [Eq.~\eqref{eq:barrier}] leading to a decrease of \TT{} and and increase of $\Delta\uptheta_F$ [Fig.~\ref{fig:Fdep}(f)], i.e. an increase of the switched fraction $N^\mathrm{SW}\propto\Delta\uptheta_F$.

\subsection{Coexistence of \Gtwo{} and \Gfour{} phases \label{sec:coexistance}}

The precession of magnetization that takes place during the switching demonstrates a peculiar character, with beatings, that are most evident when the pump photon energy is of 3.0~eV [Fig.~\ref{fig:Edep}(a)].
Such a feature was reported earlier for, e.g., the THz-pulse-driven switching in orthoferrite TmFeO$_3$ via the 2nd order SR transition \cite{schlauderer2019temporal} and was attributed to the competing contributions from magnetic precession in non-perturbative and perturbative regimes.
While the switching mechanism considered here is different and the SR transition is of the 1st order, the observed temporal behavior can also be understood along similar lines.
Indeed, two frequencies indicate that there are at least two magnetization trajectories corresponding to different potentials.
In our experimental geometry for the Sample A,  magnetization precession is detected only around the potential minimum with $\upphi=\uppi/2$  [Fig.~\ref{fig:structure}(d)].
Sample AB potentially allows detection of the precession around the initial potential minimum with $\upphi=0$. 
However, a close resemblance between the transients obtained under the same conditions for the two samples [Fig.~\ref{fig:signal}(a,b)] suggests that it is sufficient to consider only the magnetization precession around $\upphi=\uppi/2$. 

As was established in Sec.\ref{sec:Fluence}, the frequency $f_2$ agrees well with the qFMR frequency at elevated temperature in the \Gfour{} phase.
Frequency $f_1$ is lower and signifies the precession in the shallower minimum.
Based on this, we conjecture that following the thermally activated switching over the barrier, two precession trajectories are available, as schematically shown in Fig.~\ref{fig:phase}(f).
The first one corresponds to the local minimum at $\upphi=\uppi/2$ in the potential supporting \Gtwo+\Gfour{}.
The precession frequency is $f_1$.
It is defined by $D+\Delta D,~d+\Delta d$.
Accordingly, net magnetization is $\propto[D+\Updelta D-(d+\Updelta d)]$, and is lower than the equilibrium magnetization in \Gfour{} phase, as depicted in lower panels in Fig.~\ref{fig:phase}(f).
The second trajectory corresponds to the global minimum in the potential supporting only \Gfour{} phase, with equilibrium qFMR frequency $f_2$. 


Notably, the lower-frequency precession emerges at somewhat lower fluences than the higher-frequency one.
This suggests that concomitant observation of the two precession trajectories is enabled by the inhomogeneous spatial pump excitation profile.
The fact that the beating is much more pronounced under the excitation with the pump pulses with $E=3.0$~eV can be a result of different excitation pathways.
At this pump photon energy, both the change of $D,d$ and the heating are due to the one-photon absorption processes.
As a result, both $\Delta D,~\Delta d$ and of $T_\mathrm{F}$ possess similar spatial distribution.
At other pump photon energies, $\Delta D,~\Delta d$ relies of the two-photon absorption, and their spatial profile in narrower as compared to the profile of $T_\mathrm{F}$.


These features of the precessional switching 
show that laser pulse excitation brings \FBO{} from the initial \Gtwo{} phase to a state with coexisting \Gtwo$'$ and \Gfour$'$ phases.
Magnetization $m^a$ in the \Gtwo$'$ phase does not correspond to the that in the equilibrium \Gtwo{} phase.
Analogously, $m^c$ in the \Gfour$'$ phase differs from the equilibrium value as evident from the presence of the precession at $f_1$.  
Interestingly, these phases are not realized at equilibrium and can be considered as hidden ones.
At the final state, following the damping of the precession, the values of $\Updelta\uptheta_\mathrm{L}$ yield that there is \Gtwo'+\Gfour{} phase conxistance [right column in Fig.~\ref{fig:phase}(f)]. 

\section{\label{sec:Conclusion} Conclusion}

In conclusions, we have unraveled kinetics of the laser-induced spin-reorientation phase transition \Gtwo$\rightarrow$\Gfour{} in the weak ferromagnet \FBO. 
The results demonstrate that the laser-induced spin reorientation in \FBO{} cannot be described as a purely heating-driven phase transition. 
Instead, we suggest and analyze the scenario when the laser excitation first modifies the weak ferromagnetic state through a transient reduction of the canting angle with the dominant contribution originating from a change of the DMI parameter, as well as the single-ion contribution. 
This modification of these both contributions creates a transient magnetic state with a two-minimum energy potential. The reorientation proceeds through thermally-assisted crossing of the barrier separating the two orientational magnetic states \Gtwo{} and \Gfour. 
The demonstrated separation between the ultrafast preparation of the magnetic state and the subsequent switching provides a framework for understanding and controlling nonequilibrium spin-reorientation dynamics.
The dependence of the transient magnetic state on photon energy and fluence, together with the control of the switching probability and latency by temperature and magnetic field, provides several independent parameters for manipulating the reorientation kinetics. 

Our findings regarding the laser-induced change of spin canting add an important contribution to a puzzle of laser-driven change of competing DMI and isotropic exchange parameters in iron oxides exhibiting weak ferromagnetism \cite{mikhaylovskiy2015ultrafast, mikhaylovskiy2020resonant, Kuppevelt-APL2026, gavrichkov2017contribution,radovskaia2026photoengineering}, when two underlying mechanisms have been put forward based on changing the electronic configuration of Fe$^{3+}$ ions either through the CT transitions or localized $d-d$ transitions.
We demonstrate that, in the regime of strong laser-induced excitation, it is the CT transition that leads to reduced DMI and single-ion-like contributions to weak ferromagnetism by $\sim40$~\%. 
This regime is realized by combining a high base temperature of the sample and the excitation in the spectral range of low or moderate absorption.
Such conditions are readily available in a number of weak ferromagnetic iron oxides possessing ordering temperatures well above room temperature.
Finally, we note that a recent study of the related orthoferrite DyFeO$_3$ suggested a laser-induced suppression of the intersublattice isotropic exchange interaction by nearly 90~\% driven by CT transitions \cite{radovskaia2026photoengineering}, that clearly is in contrasts to our findings. 
The effect was confined to a near surface region and probed through magnon spectrum.
Thus, joint consideration of recent reports raises several interesting questions, such as whether mechanisms of laser-induced DMI and exchange perturbations strongly differ even for related iron oxides, whether surface and bulk magnetic states evolve distinctly under laser excitation, and whether different probing approaches demonstrate sensitivity to different processes.

\begin{acknowledgments}
We thank L.~A. Shelukhin for useful discussions.
We thank L.~N. Bezmaternykh for providing the Sample AB.
The work of A.V.K. and A.M.K was supported by the Russian Science Foundation (grant No.~26-12-00401).
The work of S.N.B. was supported by the Belarusian Republican Foundation for Fundamental Research (grant No.~25KI-070).
\end{acknowledgments}

\bibliography{apssamp}

@article{wolfe1969magnetization,
  title={Magnetization and {M}{\"o}ssbauer effect in single crystal {F}e$_3${BO}$_6$},
  author={Wolfe, R and Pierce, R D and Eibsch{\"u}tz, M and Nielsen, J W},
  journal={Solid State Communications},
  volume={7},
  number={13},
  pages={949--952},
  year={1969},
  publisher={Elsevier}
}

@article{tsymbal2006orientation,
  title={Orientation phase transition in {F}e$_3${BO}$_6$: Experimental determination of the order of the transition},
  author={Tsymbal, L T and Bazaliy, Ya B and Bezmaternykh, L N and Slawska-Waniewska, A and Vasiliev, S V and Nedelko, N and Linnik, A I and Cherkasov, A N and Nepochatykh, Yu I and Dmitrenko, V Yu and others},
  journal={Physical Review B},
  volume={74},
  number={13},
  pages={134429},
  year={2006},
  publisher={APS},
doi={10.1103/PhysRevB.74.134429},
url={https://doi.org/10.1103/PhysRevB.74.134429}
}

@article{kamzin1993mossbauer,
  title={M{\"o}ssbauer investigation of the surface and volume spin-reorientation phase transition in {F}e$_3${BO}$_6$},
  author={Kamzin, A S and Grigor'ev, L A},
  journal={JETP Lett},
  volume={57},
  number={9},
  pages={552--556},
  year={1993}
}

@article{kimel2004laser,
  title={Laser-induced ultrafast spin reorientation in the antiferromagnet {TmFeO}$_3$},
  author={Kimel, A V and Kirilyuk, Andrei and Tsvetkov, Artem and Pisarev, R V and Rasing, Th},
  journal={Nature},
  volume={429},
  number={6994},
  pages={850--853},
  year={2004},
  publisher={Nature Publishing Group UK London}
}

@article{weiss2023discovery,
  title={Discovery of ultrafast spontaneous spin switching in an antiferromagnet by femtosecond noise correlation spectroscopy},
  author={Weiss, M A and Herbst, A and Schlegel, Julius and Dannegger, Tobias and Evers, Martin and Donges, Andreas and Nakajima, M and Leitenstorfer, A and Goennenwein, STB and Nowak, U and others},
  journal={Nature Communications},
  volume={14},
  number={1},
  pages={7651},
  year={2023},
  publisher={Nature Publishing Group UK London}
}

@article{mikhaylovskiy2015ultrafast,
  title={Ultrafast optical modification of exchange interactions in iron oxides},
  author={Mikhaylovskiy, R V and Hendry, Euan and Secchi, A and Mentink, Johan H and Eckstein, Martin and Wu, A and Pisarev, R V and Kruglyak, V V and Katsnelson, M I and Rasing, Th and others},
  journal={Nature Communications},
  volume={6},
  number={1},
  pages={8190},
  year={2015},
  publisher={Nature Publishing Group UK London}
}

@article{ozhogin1972statics,
  title={Statics and Linear Dynamics of Orthoferrites-Phase Transition of the "One-and-One-Halfth Kind"},
  author={Ozhogin, V I and Shapiro, V G and Gurtovoi, K G and Galst'yan, E A and Chervonenkis, A Ya},
  journal={Soviet Physics JETP},
  volume={35},
pages={1162},
  number={6},
  year={1972}
}

@article{de2013speed,
  title={Speed limit of the insulator--metal transition in magnetite},
  author={de Jong, S and Kukreja, R and Trabant, Christoph and Pontius, N and Chang, C F and Kachel, T and Beye, Martin and Sorgenfrei, F and Back, C H and Br{\"a}uer, B and others},
  journal={Nature Materials},
  volume={12},
  number={10},
  pages={882--886},
  year={2013},
  publisher={Nature Publishing Group UK London}
}

@article{voigt1975temperature,
  title={Temperature dependence of the magnetic susceptibility of {F}e$_3${BO}$_6$},
  author={Voigt, C and Bonnenberg, D},
  journal={Physica B+ C},
  volume={80},
  number={1-4},
  pages={439--443},
  year={1975},
  publisher={Elsevier}
}

@article{de2011laser,
  title={Laser-induced ultrafast spin dynamics in {E}r{F}e{O}$_3$},
  author={de Jong, J A and Kimel, A V and Pisarev, R V and Kirilyuk, Andrei and Rasing, Th},
  journal={Physical Review B—Condensed Matter and Materials Physics},
  volume={84},
  number={10},
  pages={104421},
  year={2011},
  publisher={APS}
}

@article{li2022ultrafast,
  title={Ultrafast kinetics of the antiferromagnetic-ferromagnetic phase transition in {F}e{R}h},
  author={Li, G and Medapalli, R and Mentink, J H and Mikhaylovskiy, R V and Blank, TGH and Patel, S K K and Zvezdin, A K and Rasing, Th and Fullerton, E E and Kimel, A V},
  journal={Nature Communications},
  volume={13},
  number={1},
  pages={2998},
  year={2022},
  publisher={Nature Publishing Group UK London}
}

@book{gurevich2020magnetization,
  title={Magnetization oscillations and waves},
  author={Gurevich, Alexander G and Melkov, Gennadii A},
  year={2020},
  publisher={CRC press}
}

@article{andlauer1977optical,
  title={Optical Absorption of {F}e$_3${BO}$_6$ in the Range of the Transitions {$^6$A$_1\rightarrow^ 4$T$_1$, $^4$T$_2$}},
  author={Andlauer, B and Diehl, R},
  journal={Physica B+C},
  volume={89},
  pages={50--53},
  year={1977},
  publisher={Elsevier},
  doi={10.1016/0378-4363(77)90050-X} 
}

@article{belov1976spin,
  title={Spin-reorientation transitions in rare-earth magnets},
  author={Belov, Konstantin P and Zvezdin, Anatolii K and Kadomtseva, Antonina M and Levitin, Rudolf Zinovievich},
  journal={Soviet Physics Uspekhi},
  volume={19},
  number={7},
  pages={574},
  year={1976},
  publisher={IOP Publishing}
}

@article{arutyunyan1989magnetic,
  title={Magnetic resonance in the orthorhombic antiferromagnet {F}e$_3${BO}$_6$},
  author={Arutyunyan, V E and Kocharyan, K N and Martirosyan, R M},
  journal={Soviet Physics JETP},
  volume={69},
  number={4},
  pages={783},
  year={1989}
}

@article{shelukhin2022spin,
  title={Spin reorientation transition in {C}o{F}e{B}/{M}g{O}/{C}o{F}e{B} tunnel junction enabled by ultrafast laser-induced suppression of perpendicular magnetic anisotropy},
  author={Shelukhin, Leonid A and Gareev, Rashid R and Zbarsky, Vladyslav and Walowski, Jakob and M{\"u}nzenberg, Markus and Pertsev, Nikolay A and Kalashnikova, Alexandra M},
  journal={Nanoscale},
  volume={14},
  number={22},
  pages={8153--8162},
  year={2022},
  publisher={Royal Society of Chemistry}
}

@article{ichikawa2011transient,
  title={Transient photoinduced ‘hidden’ phase in a manganite},
  author={Ichikawa, Hirohiko and Nozawa, Shunsuke and Sato, Tokushi and Tomita, Ayana and Ichiyanagi, Kouhei and Chollet, Matthieu and Guerin, Laurent and Dean, Nicky and Cavalleri, Andrea and Adachi, Shin-ichi and others},
  journal={Nature Materials},
  volume={10},
  number={2},
  pages={101--105},
  year={2011},
  publisher={Nature Publishing Group UK London}
}

@article{Koshihara-PhysRep2022,
title = {Challenges for developing photo-induced phase transition {(PIPT)} systems: From classical (incoherent) to quantum (coherent) control of {PIPT} dynamics},
journal = {Physics Reports},
volume = {942},
pages = {1-61},
year = {2022},
issn = {0370-1573},
doi = {https://doi.org/10.1016/j.physrep.2021.10.003},
url = {https://www.sciencedirect.com/science/article/pii/S0370157321003744},
author = {Shinya Koshihara and Tadahiko Ishikawa and Yoichi Okimoto and Ken Onda and Ryo Fukaya and Masaki Hada and Yasuhiko Hayashi and Sumio Ishihara and Tadeusz Luty},
}

@article{kuzikova2023laser,
  title={Laser-driven first-order spin reorientation and {V}erwey phase transitions in magnetite {F}e$_3${O}$_4$ beyond the range of thermodynamic equilibrium},
  author={Kuzikova, A V and Shelukhin, L A and Maksimov, F M and Chernov, A I and Pisarev, R V and Kalashnikova, A M},
  journal={Physical Review B},
  volume={107},
  number={2},
  pages={024413},
  year={2023},
  publisher={APS}
}

@article{schlauderer2019temporal,
  title={Temporal and spectral fingerprints of ultrafast all-coherent spin switching},
  author={Schlauderer, Stefan and Lange, Christoph and Baierl, Sebastian and Ebnet, Thomas and Schmid, Christoph P and Valovcin, D C and Zvezdin, A K and Kimel, A V and Mikhaylovskiy, R V and Huber, Rupert},
  journal={Nature},
  volume={569},
  number={7756},
  pages={383--387},
  year={2019},
  publisher={Nature Publishing Group UK London}
}

@article{de2012coherent,
  title={Coherent Control of the Route of an Ultrafast Magnetic Phase Transition via Low-Amplitude Spin Precession},
  author={de Jong, J A and Razdolski, I and Kalashnikova, A M and Pisarev, R V and Balbashov, A M and Kirilyuk, Andrei and Rasing, Th and Kimel, A V},
  journal={Physical Review Letters},
  volume={108},
  number={15},
  pages={157601},
  year={2012},
  publisher={APS}
}

@article{afanasiev2021ultrafast,
  title={Ultrafast control of magnetic interactions via light-driven phonons},
  author={Afanasiev, Dmytro and Hortensius, J R and Ivanov, B A and Sasani, Alireza and Bousquet, Eric and Blanter, Y M and Mikhaylovskiy, R V and Kimel, A V and Caviglia, A D},
  journal={Nature Materials},
  volume={20},
  number={5},
  pages={607--611},
  year={2021},
  publisher={Nature Publishing Group UK London}
}

@article{Kurihara-PRL2018,
  title = {Macroscopic Magnetization Control by Symmetry Breaking of Photoinduced Spin Reorientation with Intense Terahertz Magnetic Near Field},
  author = {Kurihara, Takayuki and Watanabe, Hiroshi and Nakajima, Makoto and Karube, Shutaro and Oto, Kenichi and Otani, YoshiChika and Suemoto, Tohru},
  journal = {Phys. Rev. Lett.},
  volume = {120},
  issue = {10},
  pages = {107202},
  numpages = {6},
  year = {2018},
  month = {Mar},
  publisher = {American Physical Society},
  doi = {10.1103/PhysRevLett.120.107202},
  url = {https://link.aps.org/doi/10.1103/PhysRevLett.120.107202}
}

@article{brown1963thermal,
  title={Thermal fluctuations of a single-domain particle},
  author={Brown Jr, William Fuller},
  journal={Physical Review},
  volume={130},
  number={5},
  pages={1677},
  year={1963},
  publisher={APS}
}

@article{radu2011transient,
  title={Transient ferromagnetic-like state mediating ultrafast reversal of antiferromagnetically coupled spins},
  author={Radu, I and Vahaplar, K and Stamm, C and Kachel, T and Pontius, N and D{\"u}rr, H A and Ostler, T A and Barker, J and Evans, R F L and Chantrell, R W and others},
  journal={Nature},
  volume={472},
  number={7342},
  pages={205--208},
  year={2011},
  publisher={Nature Publishing Group UK London}
}

@article{dzyaloshinsky1958thermodynamic,
  title={A thermodynamic theory of “weak” ferromagnetism of antiferromagnetics},
  author={Dzyaloshinsky, Igor E},
  journal={Journal of Physics and Chemistry of Solids},
  volume={4},
  number={4},
  pages={241--255},
  year={1958},
  publisher={Elsevier}
}

@article{pisarev2009charge,
  title={Charge transfer transitions in multiferroic {B}i{F}e{O}$_3$ and related ferrite insulators},
  author={Pisarev, R V and Moskvin, A S and Kalashnikova, A M and Rasing, Th},
  journal={Physical Review B Condensed Matter and Materials Physics},
  volume={79},
  number={23},
  pages={235128},
  year={2009},
  publisher={APS}
}

@article{mikhaylovskiy2020resonant,
	title={Resonant pumping of d- d crystal field electronic transitions as a mechanism of ultrafast optical control of the exchange interactions in iron oxides},
	author={Mikhaylovskiy, R V and Huisman, T J and Gavrichkov, V A and Polukeev, S I and Ovchinnikov, S G and Afanasiev, D and Pisarev, R V and Rasing, Th and Kimel, A V},
	journal={Physical Review Letters},
	volume={125},
	number={15},
	pages={157201},
	year={2020},
	publisher={APS}
}

@article{moriya1960anisotropic,
	title={Anisotropic superexchange interaction and weak ferromagnetism},
	author={Moriya, T{\^o}ru},
	journal={Physical Review},
	volume={120},
	number={1},
	pages={91},
	year={1960},
	publisher={APS}
}

@article{turov1965physical,
	title={Physical properties of magnetically ordered crystals},
	author={Turov, Evgenii Akimovich and others},
journal={(No Title)},
year={1965}
}

@article{Moriya-PR1960-NiF2,
	title = {Theory of Magnetism of {Ni${\mathrm{F}}_{2}$}},
	author = {Moriya, T\^oru},
	journal = {Physical Review},
	volume = {117},
	issue = {3},
	pages = {635--647},
	numpages = {0},
	year = {1960},
	month = {Feb},
	publisher = {American Physical Society},
	doi = {10.1103/PhysRev.117.635},
	url = {https://link.aps.org/doi/10.1103/PhysRev.117.635}
}

@article{Jo-PRL2025,
	title = {Weak Ferromagnetism in Altermagnets from Alternating $g$-Tensor Anisotropy},
	author = {Jo, Daegeun and Go, Dongwook and Mokrousov, Yuriy and Oppeneer, Peter M. and Cheong, Sang-Wook and Lee, Hyun-Woo},
	journal = {Physical Review Letters},
	volume = {134},
	issue = {19},
	pages = {196703},
	numpages = {9},
	year = {2025},
	month = {May},
	publisher = {American Physical Society},
	doi = {10.1103/PhysRevLett.134.196703},
	url = {https://link.aps.org/doi/10.1103/PhysRevLett.134.196703}
}

@article{Dzyaloshinskiy-JPCS1958,
	title = {A thermodynamic theory of “weak” ferromagnetism of antiferromagnetics},
	journal = {Journal of Physics and Chemistry of Solids},
	volume = {4},
	number = {4},
	pages = {241-255},
	year = {1958},
	issn = {0022-3697},
	doi = {https://doi.org/10.1016/0022-3697(58)90076-3},
	url = {https://www.sciencedirect.com/science/article/pii/0022369758900763},
	author = {I E Dzyaloshinsky}
}

@article{Moriya-PR1960,
	title = {Anisotropic Superexchange Interaction and Weak Ferromagnetism},
	author = {Moriya, T\^oru},
	journal = {Physical Review},
	volume = {120},
	issue = {1},
	pages = {91--98},
	numpages = {0},
	year = {1960},
	month = {Oct},
	publisher = {American Physical Society},
	doi = {10.1103/PhysRev.120.91},
	url = {https://link.aps.org/doi/10.1103/PhysRev.120.91}
}

@article{Matarrese-PRB1954,
	title = {Magnetic Anisotropy of {NiF$_{2}$}},
	author = {Matarrese, L. M. and Stout, J. W.},
	journal = {Physical Review},
	volume = {94},
	issue = {6},
	pages = {1792--1793},
	numpages = {0},
	year = {1954},
	month = {Jun},
	publisher = {American Physical Society},
	doi = {10.1103/PhysRev.94.1792},
	url = {https://link.aps.org/doi/10.1103/PhysRev.94.1792}
}

@article{Dzyaloshinskiy-JETP1958-fluorides,
	title = {The magnetic structure of fluorides and the transition metals},
	author = {Dzialoshinskii, I. E. },
	journal = {Sov. Phys. - JETP},
	volume = {6},
	issue = {33},
	pages = {1120--1122},
	numpages = {0},
	year = {1958},
	month = {Jun},
	url = {https://jetp.ras.ru/cgi-bin/r/index/r/33/6/p1454?a=list}
}

@article{Moskvin-CondMat2019,
	AUTHOR = {Moskvin, Alexander},
	TITLE = {{D}zyaloshinskii–{M}oriya Coupling in 3\textit{d} Insulators},
	JOURNAL = {Condensed Matter},
	VOLUME = {4},
	NUMBER = {4},
	pages = {84},
    YEAR = {2019},
	URL = {https://www.mdpi.com/2410-3896/4/4/84},
	ISSN = {2410-3896},
	DOI = {10.3390/condmat4040084}
}

@book{Turov-book,
	author = {Turov, E. A. and Kolchanov, A. B. and Men'shenin, V. V. and Mirsaev, I. F. and Nikolaev, V. V.},
	title = {Symmetry and physical properties of antiferromagnets},
	publisher = {FizMatLit, Moscow},
	year = {2001 \textit{(in Russian)}}
}

@article{Mazurenko-JETP2021,
	AUTHOR = {Mazurenko, V. V. and Kvashnin, Y. O. and Lichtenstein, A. I. and Katsnelson, M. I.},
	TITLE = {A {D}zyaloshinskii-{M}oriya interaction guide to magnets micro-world},
	JOURNAL = {J. Exp. Theor. Phys.},
	VOLUME = {132},
	YEAR = {2021},
	NUMBER = {4},
	pages = {506},
	URL = {https://link.springer.com/article/10.1134/S1063776121040178},
	ISSN = {1063-7761},
	DOI = {10.1134/S1063776121040178}
}

@article{Kuzikova-PRL2025,
  title = {Switching of an Antiferromagnet Controlled by Spin Canting in a Laser-Induced Hidden Phase},
  author = {Kuzikova, A. V. and Liubachko, N. A. and Barilo, S. N. and Sadovnikov, A. V. and Pisarev, R. V. and Kalashnikova, A. M.},
  journal = {Physical Review Letters},
  volume = {135},
  issue = {26},
  pages = {266705},
  numpages = {9},
  year = {2025},
  month = {Dec},
  publisher = {American Physical Society},
  doi = {10.1103/bxnn-zckm},
  url = {https://link.aps.org/doi/10.1103/bxnn-zckm}
}

@article{white1965fe3bo6,
  title={Fe3BO6, a borate isostructural with the mineral norbergite},
  author={White, JG and Miller, A and Nielsen, RE},
  journal={Acta Crystallographica},
  volume={19},
  number={6},
  pages={1060--1061},
  year={1965},
  publisher={International Union of Crystallography}
}

@article{morin1950magnetic,
  title={Magnetic susceptibility of $\alpha$-{Fe$_2$O$_3$} and $\alpha$-{Fe$_2$O$_3$} with added titanium},
  author={Morin, F J},
  journal={Physical Review},
  volume={78},
  number={6},
  pages={819},
  year={1950},
  publisher={APS}
}

@article{Kuppevelt-APL2026,
    author = {van Kuppevelt, P. M. P. and Kimel, A. V.},
    title = {Laser-induced dynamics of the magnon spectrum probed by double-pump {TH}z spectroscopy},
    journal = {Applied Physics Letters},
    volume = {129},
    number = {13},
    pages = {132405},
    year = {2026},
    month = {10},
    issn = {0003-6951},
    doi = {10.1063/5.0340354},
    url = {https://doi.org/10.1063/5.0340354},
}

@article{gavrichkov2017contribution,
  title={Contribution from optically excited many-electron states to the superexchange interaction in {M}ott-{H}ubbard insulators},
  author={Gavrichkov, Vladimir A and Polukeev, Semen I and Ovchinnikov, Sergey G},
  journal={Physical {R}eview {B}},
  volume={95},
  number={14},
  pages={144424},
  year={2017},
  publisher={APS}
}

@article{radovskaia2026photoengineering,
  title={Photoengineering the magnon spectrum in an insulating antiferromagnet},
  author={Radovskaia, Viktoriia and Andrei, Radu and Hortensius, Jorrit R and Mikhaylovskiy, Rostislav V and Citro, Roberta and Chattopadhyay, Sambuddha and Na, MX and Ivanov, Boris A and Demler, Eugene and Kimel, Alexey V and others},
  journal={Nature {P}hysics},
  pages={1--8},
  year={2026},
  publisher={Nature Publishing Group UK London}
}

@article{treves1962magnetic,
  title={Magnetic studies of some orthoferrites},
  author={Treves, D},
  journal={Physical {R}eview},
  volume={125},
  number={6},
  pages={1843},
  year={1962},
  publisher={APS}
}

@article{li2026discovery,
  title={Discovery of an intrinsic non-Hermitian phase transition in a bulk condensed-matter system},
  author={Li, Jingwen and Turaev, Michael and Matsubara, Masakazu and Kliemt, Kristin and Krellner, Cornelius and Pal, Shovon and Fiebig, Manfred and Kroha, Johann},
  journal={Science},
  volume={393},
  number={6816},
  pages={1152--1155},
  year={2026},
  publisher={American Association for the Advancement of Science}
}

@article{mariager2012structural,
  title={Structural and magnetic dynamics of a laser induced phase transition in {F}e{R}h},
  author={Mariager, Simon Oddsson and Pressacco, F and Ingold, G and Caviezel, A and M{\"o}hr-Vorobeva, E and Beaud, P and Johnson, S L and Milne, C J and Mancini, E and Moyerman, S and others},
  journal={Physical {R}eview {L}etters},
  volume={108},
  number={8},
  pages={087201},
  year={2012},
  publisher={APS}
}

@article{stamps2000mechanisms,
  title={Mechanisms for exchange bias},
  author={Stamps, R L},
  journal={Journal of Physics D: Applied Physics},
  volume={33},
  number={23},
  pages={R247--R268},
  year={2000}
}

@article{Mryasov2005twoion,
doi = {10.1209/epl/i2004-10404-2},
url = {https://doi.org/10.1209/epl/i2004-10404-2},
year = {2005},
month = {feb},
publisher = {},
volume = {69},
number = {5},
pages = {805},
author = {O. N. Mryasov and U. Nowak and K. Y. Guslienko and R. W. Chantrell},
title = {Temperature-dependent magnetic properties of {F}e{P}t: 
Effective spin Hamiltonian model},
journal = {Europhysics Letters}
}

@article{zeuschner2020measurement,
  title={Measurement of transient strain induced by two-photon excitation},
  author={Zeuschner, Steffen Peer and Pudell, J-E and von Reppert, Alexander and Deb, Marwan and Popova, E and Keller, N and R{\"o}ssle, M and Herzog, Marc and Bargheer, Matias},
  journal={Physical {R}eview {R}esearch},
  volume={2},
  number={2},
  pages={022013},
  year={2020},
  publisher={APS}
}

@article{chaban2022nonlinear,
  title={Nonlinear optical absorption in nanoscale films revealed through ultrafast acoustics},
  author={Chaban, Ievgeniia and Deska, Radoslaw and Privault, Gael and Trzop, Elzbieta and Lorenc, Maciej and Kooi, Steven E and Nelson, Keith A and Samoc, Marek and Matczyszyn, Katarzyna and Pezeril, Thomas},
  journal={Nano {L}etters},
  volume={22},
  number={11},
  pages={4362--4367},
  year={2022},
  publisher={ACS Publications}
}

@article{martynov2018single,
  title={Single-ion mechanism of weak ferromagnetism and a spin-flop transition in one-and two-position antiferromagnets},
  author={Martynov, Sergei Nikolaevich},
  journal={{JETP} {L}etters},
  volume={108},
  number={3},
  pages={196--200},
  year={2018},
  publisher={Springer}
}

@article{petrakovskiui2009magnetic,
  title={Magnetic properties of {Pb$_2$Fe$_2$Ge$_2$O$_9$} single crystals},
  author={Petrakovski, G A and Popov, Michael A and Balaev, Alexandr Dmitrievich and Sablina, Klara Aleksandrovna and Bayukov, Oleg Artem'evich and Velikanov, Dmitrii Anatol'evich and Vorotynov, Alexandr Mikhailovich and Bovina, Asya Fedorovna and Vasil’ev, Aleksandr Dmitrievich and Boehm, M},
  journal={Physics of the {S}olid {S}tate},
  volume={51},
  number={9},
  pages={1853--1858},
  year={2009},
  publisher={Springer}
}

@article{WEGKAMP2015,
title = {Ultrafast dynamics during the photoinduced phase transition in VO2},
journal = {Progress in Surface Science},
volume = {90},
number = {4},
pages = {464-502},
year = {2015},
issn = {0079-6816},
doi = {https://doi.org/10.1016/j.progsurf.2015.10.001},
url = {https://www.sciencedirect.com/science/article/pii/S0079681615000301},
author = {Daniel Wegkamp and Julia Stähler}
}

@article{Fausti2011,
author = {D. Fausti  and R. I. Tobey  and N. Dean  and S. Kaiser  and A. Dienst  and M. C. Hoffmann  and S. Pyon  and T. Takayama  and H. Takagi  and A. Cavalleri },
title = {Light-Induced Superconductivity in a Stripe-Ordered Cuprate},
journal = {Science},
volume = {331},
number = {6014},
pages = {189-191},
year = {2011},
doi = {10.1126/science.1197294},
URL = {https://www.science.org/doi/abs/10.1126/science.1197294},
eprint = {https://www.science.org/doi/pdf/10.1126/science.1197294}
}

\end{document}